\documentclass[aps,prd,showpacs,nofootinbib,superscriptaddress]{revtex4}
\pdfoutput=1
\usepackage{dcolumn}
\usepackage{amsmath,amssymb,setspace,graphicx,subfigure,epsfig}
\usepackage{amsfonts}
\usepackage{latexsym}
\usepackage{epsfig}
\usepackage{color}
\usepackage{nicefrac}

\newcommand{\dis}[1]{\begin{equation}\begin{split}#1\end{split}\end{equation}}
\newcommand{\be}{\begin{equation}}
\newcommand{\ee}{\end{equation}}
\def\bea{\begin{eqnarray}}
\def\eea{\end{eqnarray}}

\newcommand{\Mp}{M_{\rm p}}

\newcommand{\G}{\Gamma}

\newcommand{\Dn}{\delta N}

\newcommand{\vp}{\varphi}
\newcommand{\vps}{\varphi_*}
\newcommand{\chis}{\chi_*}

\newcommand{\ep}{\epsilon_\vp}

\newcommand{\ec}{\epsilon_\chi}

\newcommand{\etapp}{\eta_{\vp\vp}}
\newcommand{\etapc}{\eta_{\vp\chi}}
\newcommand{\etacc}{\eta_{\chi\chi}}

\newcommand{\dtvp}{\dot{\vp}}

\newcommand{\taunl}{ \tau_{\rm NL} }
\newcommand{\gnl}{ g_{\rm NL} }

\newcommand{\fnl}{f_{\rm NL}}

\newcommand{\nz}{n_\zeta}

\newcommand{\nfnl}{n_{\fnl}}
\newcommand{\ntaunl}{n_{\taunl}}

\newcommand{\Nchi}{N_\chi}
\newcommand{\Nvp}{N_\vp}
\newcommand{\Nchichi}{N_{\chi\chi}}
\newcommand{\Nvpvp}{N_{\vp\vp}}
\newcommand{\Nchivp}{N_{\vp\chi}}
\newcommand{\Nvpvpvp}{N_{\vp\vp\vp}}

\newcommand{\Nvpvpchi}{N_{\vp\vp\chi}}
\newcommand{\Nvpchichi}{N_{\vp\chi\chi}}

\def\bkone{{\bf k_1}}
\def\bktwo{{\bf k_2}}

\def\picube{(2\pi)^3}
\newcommand{\sdelta}[1]{\!\delta^{\,3}(\mathbf{#1})}

\newcommand{\calP}{{\cal P}}

\newcommand{\Gvp}{\G_\vp}
\newcommand{\Gchi}{\G_\chi}

\begin{document}
\title{Influence of Reheating on the Trispectrum and its Scale Dependence}

\author{Godfrey Leung}
\email{ppxgl@nottingham.ac.uk}
\affiliation{School of Physics and Astronomy, University of Nottingham, University Park, Nottingham, NG7 2RD, UK}

\author{Ewan R. M. Tarrant}
\email{ppxet@nottingham.ac.uk}
\affiliation{School of Physics and Astronomy, University of Nottingham, University Park, Nottingham, NG7 2RD, UK}

\author{Christian T. Byrnes}
\email{ctb22@sussex.ac.uk}
\affiliation{Astronomy Centre, University of Sussex, Brighton, BN1 9QH, UK}

\author{Edmund J. Copeland}
\email{ed.copeland@nottingham.ac.uk}
\affiliation{School of Physics and Astronomy, University of Nottingham, University Park, Nottingham, NG7 2RD, UK}


\date{\today}

\begin{abstract}
We study the evolution of the non-linear curvature perturbation during perturbative reheating, and hence how observables evolve to their final values which we may compare against observations. Our study includes the evolution of the two trispectrum parameters, $\gnl$ and $\taunl$, as well as the scale dependence of both $\fnl$ and $\taunl$. In general the evolution is significant and must be taken into account, which means that models of multifield inflation cannot be compared to observations without specifying how the subsequent reheating takes place. If the trispectrum is large at the end of inflation, it normally remains large at the end of reheating. In the classes of models we study, it remains very hard to generate $\taunl\gg\fnl^2$, regardless of the decay rates of the fields. Similarly, for the classes of models in which $\gnl\simeq\taunl$ during slow--roll inflation, we find the relation typically remains valid during reheating. Therefore it is possible to observationally test such classes of models without specifying the parameters of reheating, even though the individual observables are sensitive to the details of reheating. It is hard to generate an observably large $\gnl$ however. The runnings, $\nfnl$ and $\ntaunl$, tend to satisfy a consistency relation $\ntaunl=(3/2)\nfnl$ regardless of the reheating timescale, but are in general too small to be observed for the class of models considered.\\

\end{abstract}

\maketitle
%%%%%%%%%%%%%%%%%%%%%%%%%%%%%%%%%%%%%%%%%%%%%%

%-------------------------------------------------------------------------------------------------------------
\section{Introduction}\label{sec:intro}

Inflation has become the leading paradigm for solving the horizon, flatness and relic problems in the Standard Hot Big Bang picture (for example, see~\cite{Guth1981Inflationary,Linde1982New,Lyth2009Primordial}) and explaining the origin of structure formation in our universe. The simplest model consists of a scalar field slowly rolling down a flat potential~\cite{Linde1982New}, resulting in an exponential expansion of spacetime. More complicated, particle theory motivated models have been studied since then. With a vast number of inflationary models in the literature, it is important to constrain and test individual ones in order to make connection with particle physics models. Recently, observational constraints on the detailed statistics of $\zeta$ have emerged as a powerful tool for testing different inflationary models. Here $\zeta$ is the gauge--invariant curvature perturbation, quantifying the perturbation in total energy density of the universe.

Primordial non--Gaussianity, as an example, opens up an extra window to constrain different inflationary models. While simple single--field models predict negligible levels of non--Gaussianity~\cite{Maldacena2003NonGaussian,Seery2005Primordial}, significant non--Gaussianity can be generated by different mechanisms, such as features in the inflaton potential~\cite{Chen2007Large}, the curvaton scenario~\cite{Chambers2010NonGaussianity,Mollerach1990Isocurvature,Lyth2002Generating,Linde2006Curvaton,Malik2006Numerical}, modulated reheating/preheating~\cite{Chambers2008Lattice,Chambers2008NonGaussianity,Kofman2003Probing,Dvali2004New,Suyama2008NonGaussianity,Byrnes2009Constraints}, and an inhomogeneous end of inflation~\cite{Lyth2005Generating}. It is also possible to generate significant non--Gaussianity during multi--field inflation~\cite{Bernardeau2002NonGaussianity,Alabidi:2006hg,Byrnes2009Large,Byrnes2008Conditions,Gao:2013hn}, for a review see~\cite{Byrnes2010Review}. For a complete review of primordial non--Gaussianity, see~\cite{Chen:2010xka}. Here we focus on local type non--Gaussianity generated in multifield models via superhorizon evolution of the curvature perturbation, $\zeta$.

As emphasised in~\cite{elliston:2011,Leung:2012ve}, $\zeta$ continues to evolve after horizon--crossing in the presence of isocurvature modes, and therefore so does its statistics. Thus it is important to take into account any superhorizon evolution up to the point where all isocurvature modes are exhausted, in order to evaluate the true model predictions for the statistics of $\zeta$ to compare against observations. Reheating, as an important part of inflationary model building which involves a transfer of energy from the inflaton to the Standard Model particles, may play a role in the evolution of $\zeta$. A number of previous works in the literature have assumed that reheating is instantaneous such that $\zeta$ becomes conserved immediately~\cite{Choi2012Primordial,Sasaki2008Multibrid,Naruko2009Large}. This assumption is unrealistic in general as reheating presumably takes finite time to complete.

Peterson et.al.~\cite{Peterson:2010mv} have found compact relations between $\fnl$, $\taunl$ and $\gnl$ and the tilts of the curvature and isocurvature power spectra, under the slow--roll and slow--turn approximation. In particular, they found that for detectable non--Gaussianity in multifield models without excessive fine-tuning, $\taunl$ can not be much larger than $\fnl^2$. These results have also been verified by Elliston et.al.~\cite{Elliston2012Large} making use of explicit analytic expressions for models with separable potentials. Even when reheating is taken into account, we will find that their results continue to hold.

Using the finite central difference method, we have numerically implemented the $\Dn$ formalism to follow the evolution of $\zeta$ and its statistics for two--field models through a phase of perturbative reheating. In a previous paper~\cite{Leung:2012ve}, we have demonstrated that $\fnl$ is in general sensitive to the reheating timescale, whilst the spectral index $\nz$ is less sensitive, and therefore can be considered a better probe of the underlying inflationary model. In general, the sensitivity is model--dependent, meaning that single values of $\fnl$ can only be reliably used to discriminate between different multifield models if the physics of reheating is properly accounted for. Here we extend our previous work to the trispectrum and the running of the non--linear parameters $\nfnl$ and $\ntaunl$. Although the extension is straightforward to imagine, it is a non-trivial exercise to show that the conclusions found previously for the bispectrum apply also for trispectrum.

Besides investigating how reheating changes the inflationary predictions of these individual observables, we also explore the possible relations between different observables. The aim is to find whether there are consistency relations between observables that survive through reheating, which would therefore be a smoking gun of the scenario which gives rise to the relation. 

The results are briefly summarised as follows: As in the case for fNL~\cite{Leung:2012ve}, we find that the non--linear parameters in the trispectrum are also sensitive to the reheating timescale, while $\gnl$ remains too small to be observed in general.  The runnings $\nfnl$ and $\ntaunl$ are small except in some cases of the quadratic times exponential model. Though individual observables depend upon the reheating timescale, certain consistency relations between them in some classes of multifield models survive through reheating, offering hope to test such models without specifying the details of reheating. This is one of the main results of this paper.

The paper is organised as follows: in Section~(\ref{sec:backgroundtheory}) and Section~(\ref{sec:zeta}) we introduce some background material including the definitions of the non--linear parameters $\fnl$, $\taunl$ and $\gnl$ and the runnings $\nfnl$ and $\ntaunl$. In Section~(\ref{sec:deltaN}) we recall the $\delta N$ formalism and give the formulae of the primordial observables in terms of the $\Dn$ derivatives. Then, in Section~(\ref{sec:reheat}) we study how reheating may alter the inflationary predictions for $\taunl$, $\gnl$, $\nfnl$ and $\ntaunl$, including relations between them, in canonical two--field models. We study two classes of models where a minimum exists in one or both field directions in Section~(\ref{sec:onemin}) and~(\ref{sec:twomins}) respectively. Our discussion and conclusions are presented in Sections~(\ref{sec:disc}) and~(\ref{sec:conclus}).

The reader who is interested in the details of our numerical recipe is referred to our previous paper~\cite{Leung:2012ve}. Although we have used the same basic recipe as in our earlier work, extending the calculations to third-order in $\Dn$ derivatives is by no means a straightforward exercise. In particular, for cross derivatives terms such as $\Nvpvpchi$ and $\Nvpchichi$, we require different finite step-sizes for the initial conditions $\{\vps,\chis\}$ of the bundle of trajectories, i.e. $\delta\vps\neq\delta\chis$, in order to ensure that $\Nvpvpchi$ and $\Nvpchichi$ converge with respect to the step-sizes used. A larger grid of $\{\vps,\chis\}$ may also be needed in the numerical analysis to achieve the same accuracy in evaluating the trispectrum as in bispectrum. Moreover, deriving the required formulae for these third-order cross derivatives using the central finite difference method is an involved process that requires a great deal of care.

%-----------------------------------------------------------------------------------------------------------------
\section{Background Theory}\label{sec:backgroundtheory}

The class of two--field models considered in this paper are described by the inflationary action:
\dis{ 
\label{eq:action}
S=\int d^4 x \sqrt{-g}\left[\Mp^2\frac{R}{2}-\frac12g^{\mu\nu}\partial_\mu\vp\partial_\nu\vp-\frac12g^{\mu\nu}\partial_{\mu}\chi\partial_\nu\chi-W(\vp,\chi) \right]\,,
}
where $\Mp=1/\sqrt{8\pi G}$ is the reduced Planck mass. The standard slow-roll parameters are defined as
\bea
\label{eq:slowrollpar}
\ep = \frac{\Mp^2}{2}\left(\frac{W_{\vp}}{W}\right)^{2}\,, \quad \ec = \frac{\Mp^2}{2}\left(\frac{W_{\chi}}{W}\right)^{2}\,, \quad \epsilon = \ep + \ec \,,\nonumber \\
\etapp=\Mp^2\frac{W_{\vp\vp}}{W}\,, \quad \etapc=\Mp^2\frac{W_{\vp\chi}}{W}\,, \quad \etacc=\Mp^2\frac{W_{\chi\chi}}{W}\,,
\eea
where subscripts denote differentiation with respect to the fields, $\vp_I$.  The dynamics of the scalar fields are governed by the Klein--Gordon equation
\be 
\label{eq:KGeqn}
\ddot{\vp_{I}}+3H\dtvp_{I}+W_{\vp_{I}} =0\,,
\ee
where the first term can be neglected during slow--roll inflation. After inflation ends, the fields start to oscillate about their minima. If the period of an oscillation is much shorter than the Hubble time, the fields can be interpreted as a collection of particles with zero momenta that decay perturbatively to bosons $\chi_b$ and fermions $\psi_f$ via interaction terms like $-\frac12g^2\vp_{I}^2\chi_b^2$ and $-h\bar{\psi_f}\psi_f\vp_{I}^2$. This is the process of perturbative reheating~\cite{Kofman1996Origin}. 

As a phenomenological prescription, we model reheating by adding friction terms $\Gamma_I \dot{\vp}_I$ to the Klein-Gordon equation Eq.~(\ref{eq:KGeqn}) \cite{Leung:2012ve} 
\bea 
\label{eq:KGpheno}
\ddot{\vp_I}+(3H+\Gamma_I)\dot{\vp}_I+W_{\vp_{I}} &=&0\,,\\
\label{eq:radiation}
\dot{\rho}_\gamma+4H\rho_\gamma&=&\sum_I\Gamma_I\dot{\vp}_I^2\,,
\eea
which couples the scalar fields to an effective radiation fluid $\rho_\gamma$. The decay terms are `switched on' only when the scalar fields pass through their minima for the first time, with the conditions $m_{\vp _{I}}\gg {\rm max}\{H,\Gamma\}$ satisfied, where $\Gamma=\sum_I\Gamma_I$~\cite{Kofman1996Origin}. For simplicity, we assume the decay rates $\Gamma_I$ are constant and the decay products are relativistic and thermalised instantaneously. The completion of reheating is taken to be the time when the universe becomes radiation dominated, i.e. $\Omega_{\gamma} \sim 1$. At that point, isocurvature modes have decayed away and become negligible, hence observables freeze into their final values. Very recently a study of the possible survival of an isotropic pressure perturbation during reheating has been studied \cite{Huston:2013kgl}, where the fields are allowed to decay into both radiation and matter. Even allowing for this, in all models studied the isocurvature mode quickly becomes negligible.

%-------------------------------------------------------------------------------------------------------------
%-------------------------------------------------------------------------------------------------------------
\section{The Curvature Perturbation, $\zeta$}\label{sec:zeta}

At leading order, the statistics of $\zeta$ are measured in terms of the power spectrum, bispectrum and trispectrum, which are defined in Fourier space by
\begin{eqnarray}
\label{eq:powerspectrumdefn} 
\langle\zeta_{\bkone}\zeta_{\bktwo}\rangle &\equiv&
\picube\,
\sdelta{\bkone+\bktwo}\frac{2\pi^2}{k_1^3}\calP_{\zeta}(k_1) \, , \\
\label{eq:bispectrumdefn}
\langle\zeta_{{\mathbf k_1}}\,\zeta_{{\mathbf k_2}}\,
\zeta_{{\mathbf k_3}}\rangle &\equiv& \picube\, \sdelta{{\mathbf
k_1}+{\mathbf k_2}+{\mathbf k_3}} B_\zeta( k_1,k_2,k_3) \,, \\
\label{eq:trispectrumdefn}
\langle\zeta_{{\mathbf k_1}}\,\zeta_{{\mathbf k_2}}\,
\zeta_{{\mathbf k_3}}\,\zeta_{{\mathbf k_4}}\rangle &\equiv& \picube\, \sdelta{{\mathbf
k_1}+{\mathbf k_2}+{\mathbf k_3}+{\mathbf k_4}} T_\zeta( k_1,k_2,k_3,k_4,k_{12},k_{13}) \,.
\end{eqnarray}
where $k_{ij}\equiv |{\mathbf k_i} + {\mathbf k_j}|$. Here the delta functions are present due to the assumption of statistical homogeneity and isotropy. The level of non--Gaussianity can then be parametrized by the dimensionless non--linear parameters $\fnl$, $\taunl$ and $\gnl$,  
\begin{eqnarray}
\label{eq:fnl_defn} 
B_\zeta( k_1,k_2,k_3) &=& \frac{6}{5}\fnl\left[P_\zeta (k_1) P_\zeta(k_2) + 2\, perms \right]\,, \\
\label{eq:gnl_taunl_defn}
T_\zeta( k_1,k_2,k_3,k_4,k_{12},k_{13}) &=& \taunl\left[P_\zeta (k_{12}) P_\zeta(k_1) P_\zeta(k_3)+ 11 \,perms \right] + \frac{54}{25}\gnl\left[P_\zeta (k_1) P_\zeta(k_2) P_\zeta(k_3) + 3\, perms \right]\,,
\end{eqnarray}
which are in general functions of wavevectors $\mathbf{k_i}$ and thus are shape dependent. For canonical models however, non--Gaussianity peaks in the local shape, which is defined by $\zeta=\zeta_G + 3\fnl\zeta_G^{2}/5 + 9\gnl\zeta_G^3$ where $\zeta_G$ is the Gaussian part of $\zeta$. Here we focus on the local shape only, for which the current constraint from WMAP 9--year data on $\fnl$ is: $-3 < \fnl < 77$ at $95\%\,CL$~\cite{Bennett:2012fp}. An even tighter constraint comes from large scale structure $-37 < \fnl < 25$ at $95\%\,CL$~\cite{Giannantonio:2013uqa}. For the trispectrum, WMAP 5--year data gives the following constraints:  $-0.6 < \taunl /10^4 < 3.3$ and $-7.4 < \gnl /10^5 < 8.2$~\cite{Smidt:2010sv,Smidt:2010ra}, with~\cite{Fergusson:2010gn} finding a compatible constraint $-5.4 < \gnl /10^5 < 8.6$. A slightly tighter constraint for $\gnl$ was also found in \cite{Giannantonio:2013uqa}, but with the caveat of how to model the scale dependent bias due to $\gnl$. All of these will be improved considerably by Planck data very soon. In the absence of a detection, the bounds are given by $|\fnl|<5$~\cite{Komatsu:2001rj}, $\taunl<560$~\cite{Kogo:2006kh}, and $|\gnl|<1.6\times 10^5$~\cite{Smidt:2010ra}.\\
 
Like the power spectrum, it is natural that the non-linearity parameters are scale dependent~\cite{Chen:2005fe,Byrnes2009Large,Byrnes:2010ft,Shandera:2010ei}, quantified by their runnings. For instance, the runnings of $\fnl$ and $\taunl$, denoted by $\nfnl$ and $\ntaunl$, are defined by
\begin{eqnarray}
n_{\fnl} \equiv \frac{{\rm d\, ln}|\fnl|}{{\rm d\, ln}k}  \label{eq:defn_nfnl}\,, \\
n_{\taunl} \equiv \frac{{\rm d\, ln}|\taunl|}{{\rm d\, ln}k} \label{eq:defn_ntaunl}\,,
\end{eqnarray}
where $k$ marks the length of any one side of the $n$-gon, provided that all sides are scaled in the same proportion~\cite{Byrnes:2010ft}. Examples of models where $\nfnl$ and $\ntaunl$ can be observationally large, i.e. $O(0.1)$, are the curvaton models with quartic self-interaction terms~\cite{Byrnes:2011gh,Kobayashi:2012ba} and modulated reheating~\cite{Byrnes:2010ft}. Forecasts have been made to assess our ability detect the running of these non--linear parameters. For $\nfnl$, Planck could reach a $1-\sigma$ sensitivity of $\sigma_{\nfnl}\sim 0.1$ given $\fnl=50$~\cite{Sefusatti:2009xu}. By measurements of the CMB $\mu$-distortion in a CMB experiment such as PIXIE, $\fnl$ and $\ntaunl$ could also be measured to an accurancy of the order of $O(0.3)$ and $O(0.6)$ respectively for $\fnl = 20$ and $\taunl = 5000$, and similarly in large-scale surveys such as Euclid~\cite{Biagetti:2013sr}.\footnote{Their definition of $\ntaunl$ differs from the one used here, in the fact that in their case, $\ntaunl\neq d\ln|\taunl|/(d\ln k)$. The two definitions are related when the four $k$ vectors form a square by $2\ntaunl^{\rm theirs}=\ntaunl^{\rm ours}$, in which case we have to double their forecasted error bars when comparing to our definition of $\ntaunl$.} For any non-linearity parameter, the error bar on its scale dependence is approximately inversely proportional to its fiducial value~\cite{Sefusatti:2009xu}.

%-------------------------------------------------------------------------------------------------------------
\section{The $\delta$N Formalism}\label{sec:deltaN}

The $\delta$N formalism~\cite{Sasaki1996General,Sasaki1998SuperHorizon,Lyth2005Inflationary} has been used extensively throughout the literature to compute the primordial curvature perturbation and its statistics. The formalism relates $\zeta$ to the difference in the number of $e$--folds of expansion, $\delta N$, between different superhorizon patches of the universe, given by \cite{Lyth2005Inflationary} (or~\cite{Saffin2012Covariance,Elliston:2012ab} for the covariant approach)
\be
\label{eq:deltaN} 
\zeta=\delta N= 
N_{I}\delta\vp_{I*}+\frac{1}{2}N_{IJ}\delta\vp_{I*}\delta\vp_{J*}+\cdots\,, 
\ee
where $N$ is defined as the number of $e$--folds of expansion from an initial flat hypersurface to a final uniform energy density hypersurface. We take the initial time to be Hubble exit during inflation, denoted by $t_*$, and the final time, denoted by $t_c$, to be a time deep in the radiation dominated era when reheating has completed. All repeated indices are implicitly summed over unless stated otherwise. Here $N_I=\partial N/(\partial \vp_{I*})$, the index $I$ runs over all of the fields, and subscript $*$ denotes the values evaluated at horizon--crossing. In general, $N(t_c,t_*)$ depends on the fields, $\vp_I(t)$, and their time derivatives, $\dot{\vp}_I(t)$. However, if the slow--roll conditions, $3H\dot{\vp}_I \simeq - W_{,I}$, are satisfied at Hubble exit, then $N$ depends only on the initial field values. The radiation fluid remains effectively unperturbed at horizon exit as it does not yet exist, and so does not feature in the above expansion.

For canonical models, the non--linear parameters defined in Eq.~(\ref{eq:fnl_defn}-\ref{eq:gnl_taunl_defn}) are dominated by their shape independent parts, which under the $\Dn$ formalism are expressed as \cite{Lyth2005Inflationary,Byrnes:2006vq} 
\begin{eqnarray}
\label{eq:fnl_Dn} 
\fnl &=&\frac56 \frac{N_{IJ}N_IN_J}{( N_KN_K)^2}\,, \\
\taunl &=& \frac{N_{IJ} N_{JK}N_K N_I}{(N_LN_L)^3} \label{eq:taunl_Dn} \,,\\
\gnl &=& \frac{25}{54}\frac{N_{IJK} N_IN_JN_K}{(N_LN_L)^3}\,,
  \label{eq:gnl_Dn}
\end{eqnarray}
Similarly, in terms of $\Dn$ derivatives, the runnings $\nfnl$ and $\ntaunl$ are given by \cite{Byrnes:2010ft,Byrnes:2009pe} 
\begin{eqnarray}
\nfnl = && -2(\nz - 1 +2\epsilon_*)+\frac{5}{6\fnl}\left(\frac{1}{H_{*}}\right)\left[\frac{N_{IJK}N_{I}N_{J}(\dtvp_{K})_{*}}{(N_LN_L)^2} + 2\frac{N_{IJ}N_{IK}N_{J}(\dtvp_{K})_{*}}{(N_LN_L)^2}\right] \,, \label{eq:nfnl_Dn} \\
\ntaunl = &&-3(\nz - 1 +2\epsilon_*)+\frac{2}{\taunl}\left(\frac{1}{H_{*}}\right)\left[\frac{N_{IJL}N_{IK}N_{J}N_{K}(\dtvp_{L})_{*}}{(N_MN_M)^3} + \frac{N_{IJ}N_{IK}N_{JL}N_{K}(\dtvp_{L})_{*}}{(N_MN_M)^3}\right]\,, \label{eq:ntaunl_Dn}
\end{eqnarray}
assuming slow--roll at horizon--crossing such that $\frac{\rm d}{{\rm d\,ln}k}\approx \frac{\dtvp_{I*}}{H_*}\frac{\partial}{\partial\vp_{I*}}$. Using $\frac{{\rm d}N}{{\rm d}t_*}=-H_*$ and the slow--roll field equations, we have
\bea
N_{I}W_{I*} &=& W_{*}  \label{eq:Dn_pot_1st}\,, \\
N_{IJ}W_{I*} &=& W_{J*} - N_{I}W_{IJ*}\,,  \label{eq:Dn_pot_2nd} \\
N_{IJK}W_{I*} &=& W_{JK*} - N_{IJ}W_{IK*} - N_{IK}W_{IJ*} - N_{I}W_{IJK*} \,, \label{eq:Dn_pot_3rd}
\eea 
where Eqs.~(\ref{eq:Dn_pot_2nd}-\ref{eq:Dn_pot_3rd}) are derived by differentiating Eq.~(\ref{eq:Dn_pot_1st}) with respect to $\vp_{I}^{*}$. The results that during slow--roll, higher order $\Dn$ derivatives can be eliminated in favour of lower order ones whenever they come in combinations with $\dtvp_{I*}$ such as $N_{I}\dtvp_{I*}, N_{IJ}\dtvp_{I*}$ was first noticed by Lyth and Riotto, for instance see Eqs.~(113) and (114) in~\cite{Lyth:1998xn}, where they have used these to work out alternative expressions for $\nz$ and its running. Following a similar approach here, we extend it to the case of $\nfnl$ and $\ntaunl$. This allows us to rewrite $\nfnl$ and $\ntaunl$ in terms of only first and second-order derivatives of $N$ as follows
\bea
n_{\fnl} = && -2(\nz - 1 +2\epsilon_*) - \frac{10}{6\fnl}\left(\frac{1}{N_LN_L}\right)^2 + \frac{5}{6\fnl}\left[\frac{4\eta_{IK*}N_{IJ}N_{J}N_{K} + \eta_{IJ*}N_{I}N_{J} + (W_{IJK}/W)_*N_{I}N_{J}N_{K}}{(N_LN_L)^2}\right] \,,
\label{eq:nfnl_pot} \nonumber \\
\\
n_{\taunl} = && -3(\nz - 1 +2\epsilon_*) - \frac{2}{\taunl}\left(\frac{1}{N_MN_M}\right)^{3} \nonumber \\ 
&& + \frac{2}{\taunl}\left[\frac{2\eta_{JL*}N_{IJ}N_{IK}N_{L}N_{K} + \eta_{IJ*}N_{I}N_{J} + \eta_{IJ*}N_{JL}N_{IK}N_{L}N_{K}+(W_{IJL}/W)_*N_{IK}N_{J}N_{K}N_{L}}{(N_MN_M)^3}\right] \,.
\label{eq:ntaunl_pot}
\eea
Here $\nz - 1 $ is the spectral index. Eqs.~(\ref{eq:nfnl_pot}-\ref{eq:ntaunl_pot}) are two useful results of this paper. Whilst Eqs.~(\ref{eq:nfnl_pot}-\ref{eq:ntaunl_pot}) are equivalent to Eqs.~(\ref{eq:nfnl_Dn}-\ref{eq:ntaunl_Dn}), they possess significant computational advantages over the former since they involve lower order $\Dn$ derivatives which are relatively easier to compute in general compared to higher order ones.

%-------------------------------------------------------------------------------------------------------------
\section{Sensitivity to Reheating}\label{sec:reheat}

In this section we present numerical results for the evolution of the statistics of $\zeta$ for the class of two--field models where a minimum exists in one or both field directions, focusing on those models which can produce large values of $\fnl$ and $\taunl$ during perturbative reheating. We focus on the trispectrum, the running of $\fnl$ and $\taunl$ and consistency relations between observables. In what follows, $\chi$ may be identified as the inflaton and $\vp$ as the field which sources the isocurvature perturbations. For the one minimum case, the $\vp$ field is not directly involved in the reheating phase and so $\Gvp=0$ at all times. For models with two minima, both fields can decay to radiation and so both $\Gchi$ and $\Gvp$ can be non--zero.

%-------------------------------------------------------------------------------------------------------------
\subsection{One minimum}\label{sec:onemin}

First we consider a two--field model where a minimum exists in only one of the field directions. In particular, we study the `runaway' type model 
\be
W(\vp,\chi)=W_0\chi^2e^{-\lambda\vp^2/\Mp^2} \,. \label{eq:pot_onemin_quadexp}
\ee
This model was first introduced in~\cite{Byrnes2008Conditions} and has been studied extensively in the literature since then~\cite{elliston:2011,Dias2012Transport,Huston2012Calculating,Anderson:2012em,Watanabe2012Delta,Peterson2011NonGaussianity}. It does not possess a focussing region where the bundle of inflationary trajectories may converge, meaning the isocurvature mode would never be exhausted unless reheating is taken into account. By placing the $\vp$ field close to the top of the ridge at horizon--crossing, a large negative $\fnl$ can be produced~\cite{Byrnes2008Conditions}. Here we restrict ourselves to the parameter space where a detectable level of non--gaussianity, $|\fnl|>O(1)$, is generated by the end of inflation. \footnote{We have also studied two slightly different models where significant terms beyond quadratic order are present. The potentials are $W=W_0\chi^4e^{-\lambda\vp^2/\Mp}$ and $W=W_0\chi^2e^{-\lambda\vp^4/\Mp}$. The qualitative behaviour for these models is similar, with $\gnl$ negligible in general.}

\subsubsection{Trispectrum}

Before studying how the non--linear parameters $\taunl$ and $\gnl$ evolve during reheating, it is useful to consider their evolution during the inflationary phase. Because the potential is of product--separable form, analytic expressions exist for $\taunl$ and $\gnl$ during slow--roll. These have been studied extensively in~\cite{Elliston2012Large}. Anderson et.~al.~\cite{Anderson:2012em} have also studied the evolution of $\taunl$ and $\gnl$ in this model using the moment transport equations developed in~\cite{Mulryne2011Moment}. To summarise, a large $\taunl$ is produced in similar regions of parameter space as that of a large $\fnl$. $\gnl$ remains subdominant throughout inflation except possibly if there are significant terms beyond quadratic order in the potential.

Here we are interested in the post--inflationary evolution during reheating. In particular we study how $\taunl$ and $\gnl$ evolve with different decay rates $\Gchi$, and how their final values at the end of reheating depend on $\Gchi$. We start with $\taunl$. 
\begin{figure*}[!htb]
	\begin{tabular}{cc}
       	                \includegraphics[width=0.5\linewidth]{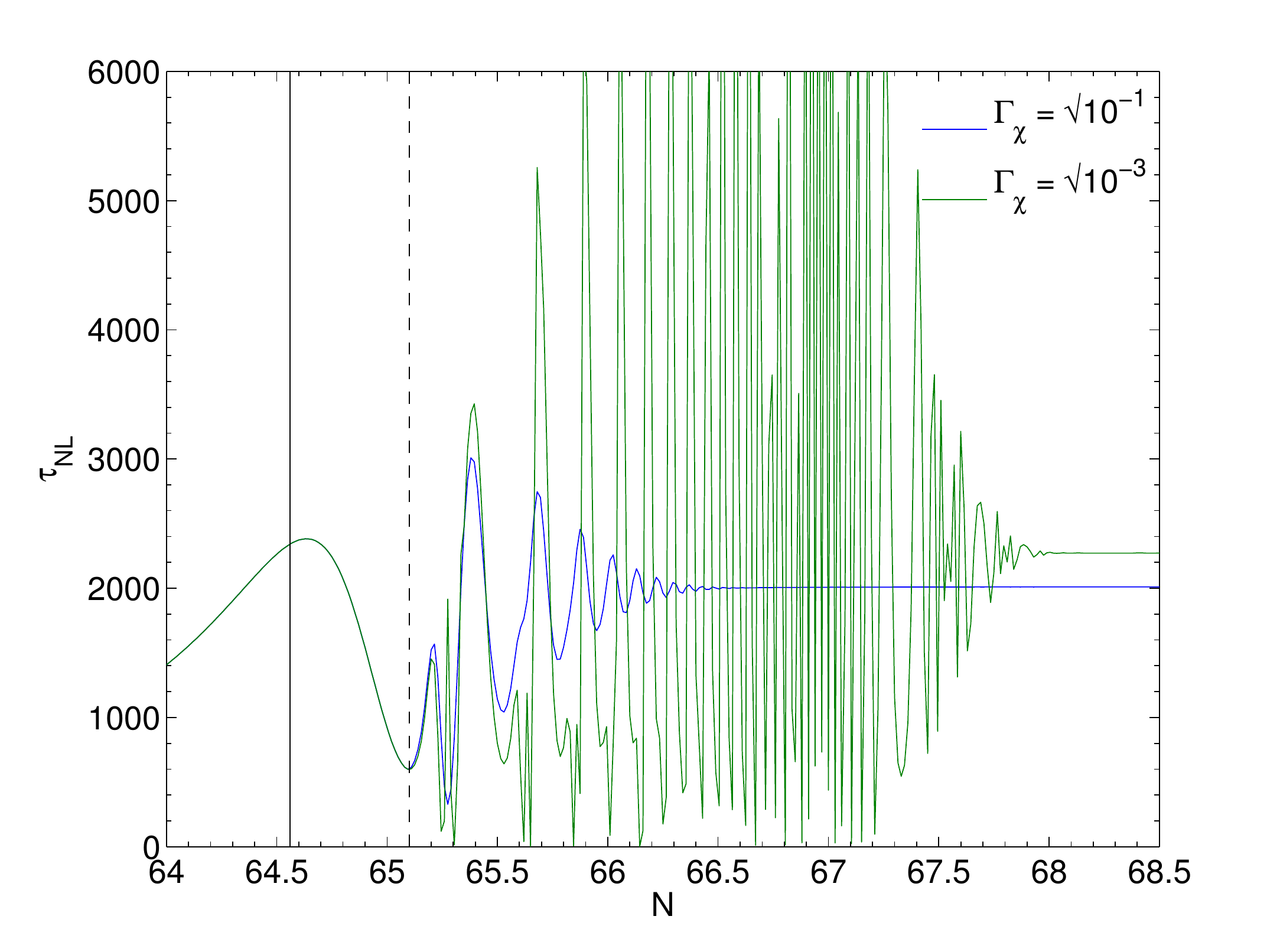}
			\includegraphics[width=0.55\linewidth]{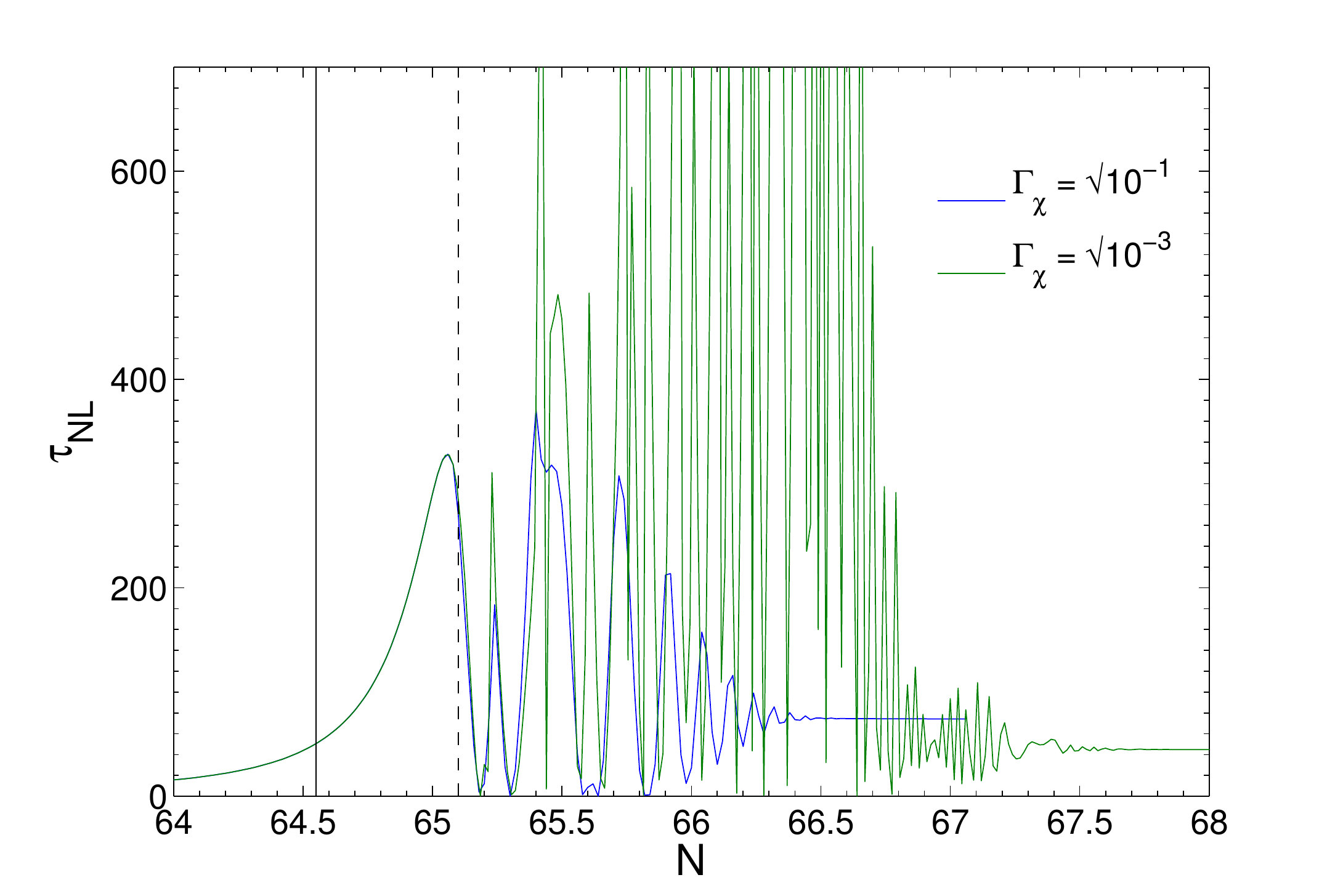}
\end{tabular}	
	\caption{Potential: $W(\vp,\chi)=W_0\chi^2e^{-\lambda\vp^2/\Mp^2}$. \textit{Left panel}: The evolution of $\tau_{NL}$ during the post--inflationary period, with $\lambda=0.05$, $\vps=10^{-3}\Mp$ and $\chis=16.0\Mp$. \textit{Right panel}: Same initial conditions with $\lambda=0.06$. The solid vertical line denotes the end of inflation, $N_{\rm e}$, and the dashed line denotes the start of reheating, $N|_{\rm\chi=0}$ here and in all subsequent figures for this one minimum model. All decay rates in this paper are given in unit of $\sqrt{W_0}\Mp$. Note the final value of $\taunl$ can either grow or decay with larger $\Gchi$, and are different from the end of inflation value.}
	\label{fig:tauNL_quadexp_reheat}
\end{figure*}
In Fig.~\ref{fig:tauNL_quadexp_reheat} we show the evolution of $\taunl$ during reheating for two different decay rates $\Gchi$, in the case of two slightly different slopes of the ridge in the potential as determined by $\lambda$. The model parameters are $\lambda=\left\{0.05,\,0.06\right\}$, $\vps=10^{-3}\Mp$ and $\chis=16\Mp$. As we found with $\fnl$ in~\cite{Leung:2012ve}, $\taunl$ oscillates during reheating when $\chi$ oscillates about its minimum. No generic trend independent of $\lambda$ can be seen as the decay rate is increased, as we see that the final value of $\taunl$ can either grow or decay as the reheating timescale increases. This may be understood by making approximations in Eq.~(\ref{eq:taunl_Dn}) and determining how the $\Dn$ derivatives evolve as follows: 

As we demonstrated in~\cite{Leung:2012ve}, during reheating the $\Nchichi$ and $\Nchivp$ are negligible compared to $\Nvpvp$. Together with the scaling relation found between $\Nvpvp$ and $\Nvp$, where $\Nvpvp\approx\Nvp/\vps$~\cite{Leung:2012ve}, we may then write $\taunl$ as
\be
   \taunl = \frac{(\Nvp^4)}{(\Nvp^2+g_{*}^2)^3}\left(\frac{1}{\vps^2}\right) \,.
\label{eq:taunl_approx_quadexp}
\ee
Here $g_{*}\equiv\Nchi\simeq (2\epsilon^{*}_{\chi})^{-1/2}$, which for this potential, is approximately constant and independent of $\lambda$. The result that $g_{*}\simeq const$ comes from the fact that the $\chi$ field dominates the energy density over the whole evolution. This algebraic function has three stationary points at certain values of $\Nvp$, 
\be
   \Nvp=0,\pm \sqrt{2}g_*\,.  \label{eq:stationarypt}
\ee
The $\Nvp=0$ root is a global minimum where $\taunl=0$, while the $\Nvp=\pm\sqrt{2}g_*$ corresponds to a maximum. Both $\Nvp=0$ and $\Nvp=\sqrt{2}g_*$ roots are unphysical here because $\Nvp$ is always negative with diverging trajectories. The other root, $\Nvp=-\sqrt{2}g_*$, however is physical and bounds the maximum value of $\taunl$, given by
\be
  (\taunl)_{max} = \frac{4}{27g_{*}^2}\left(\frac{1}{\vps^2}\right)\,.  \label{eq:taunl_max_quadexp}
\ee
This bound depends entirely on the initial conditions at horizon--crossing, not on any superhorizon evolution including reheating. The final value of $\taunl$ at the end of reheating is of course dependent on $\Gchi$ however. But since a bound exists, even if the details of reheating such as $\Gchi$ are unknown, we are still able to constrain the possible range where $\taunl$ could lie in this model.  

\begin{figure}
\begin{tabular}{cc}
		\includegraphics[width=0.5\linewidth]{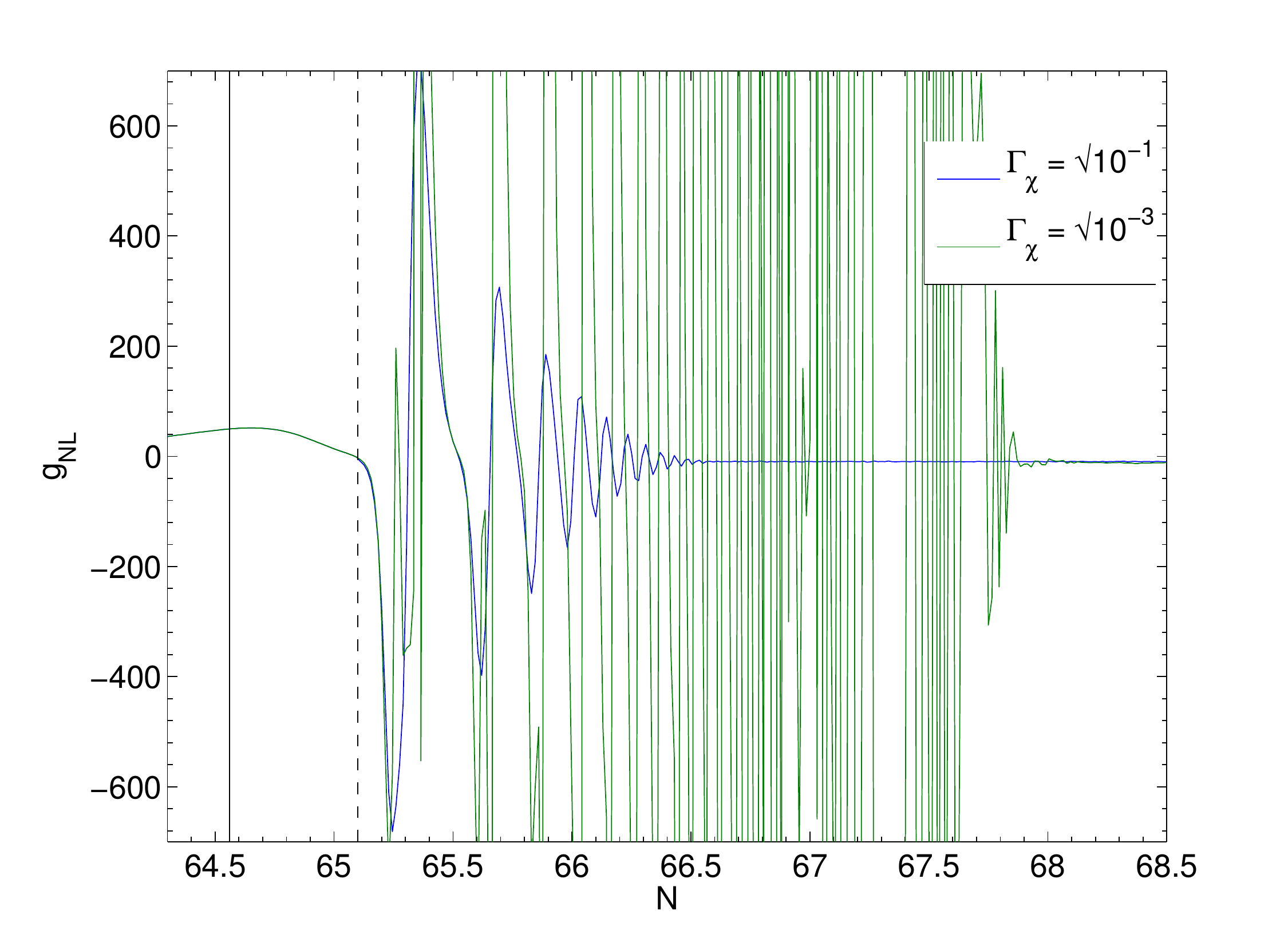}
		\includegraphics[width=0.5\linewidth]{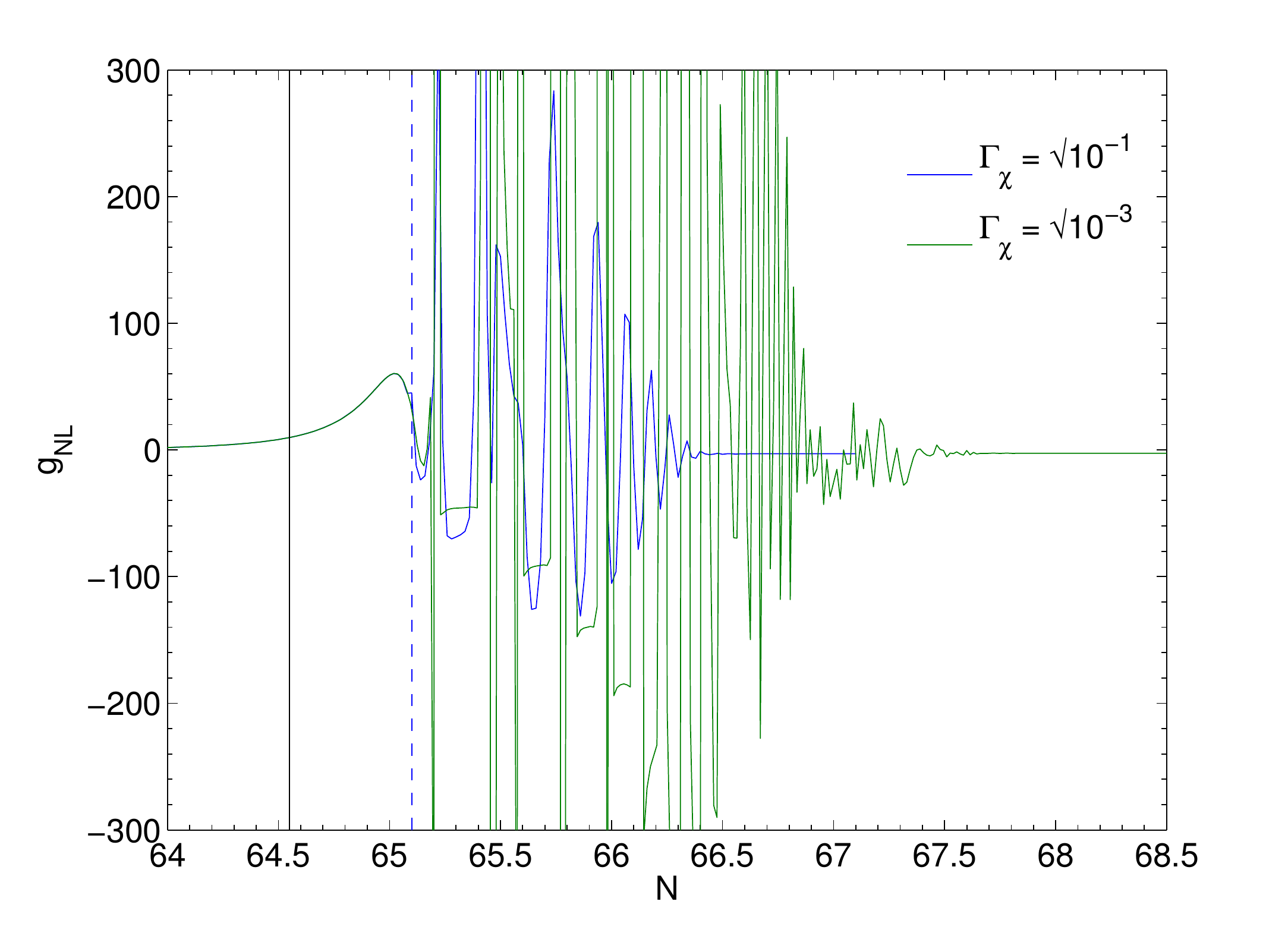}
	\end{tabular}	
	\caption{Potential: $W(\vp,\chi)=W_0\chi^2e^{-\lambda\vp^2/\Mp^2}$. The post--inflationary evolution of $\gnl$ for two different decay rate $\Gchi$. The initial conditions are $\vps=10^{-3}\Mp$ and $\chis=16.0\Mp$. \textit{Left Panel}: $\lambda=0.05$, \textit{Right Panel}: $\lambda=0.06$. Compared to $\taunl$, $\gnl$ remains subdominate after reheating and is consistent with being zero in future experiments, though it differs slightly from the end of inflation value as shown in Table~\ref{tab:oneminstats}. }
	\label{fig:gNL_quadexp}
\end{figure}

We now repeat the same analysis for $\gnl$. In Fig.~\ref{fig:gNL_quadexp} we show the evolution of $\gnl$ during reheating for two different $\Gchi$. The model parameters are $\lambda={0.05,0.06}$, $\vps=10^{-3}\Mp$ and $\chi=16\Mp$. Similar to $\fnl$ and $\taunl$, we find that $\gnl$ oscillates during reheating, with the final value at the end of reheating sensitive to the decay rate $\Gchi$.
Just as for the case of the second-order $\Dn$ derivatives, we have found that there is also a hierachy for the third-order $\Dn$ derivatives with $\Nvpvpvp$ being much larger than the other third-order derivatives. Using this, Eq.~(\ref{eq:gnl_Dn}) can be reduced to 
\be
  \gnl \approx \frac{25}{54}\frac{\Nvpvpvp\Nvp^3}{(\Nvp^2 + \Nchi^2)^3}\,.  \label{eq:gnl_quadexp_approx}
\ee
Compared to $\taunl$, however, it remains subdominate ($\ll O(10^5)$) and too small be observed in ongoing CMB experiments. 

\subsubsection{Runnings of non--linear parameters, $\nfnl$ and $\ntaunl$}

\begin{figure*}[t]
	\begin{tabular}{cc}
		\includegraphics[width=0.45\linewidth]{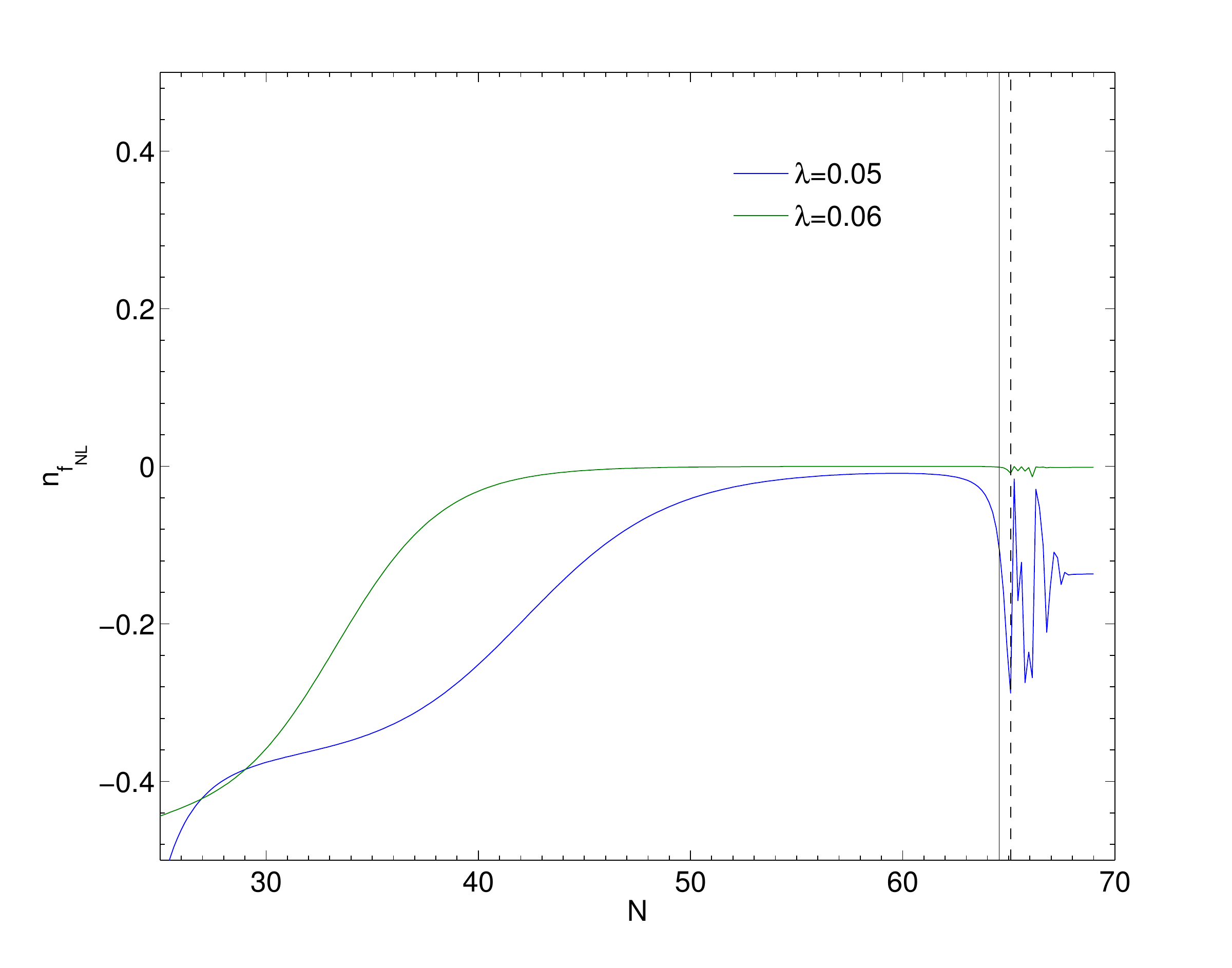}
		\includegraphics[width=0.45\linewidth]{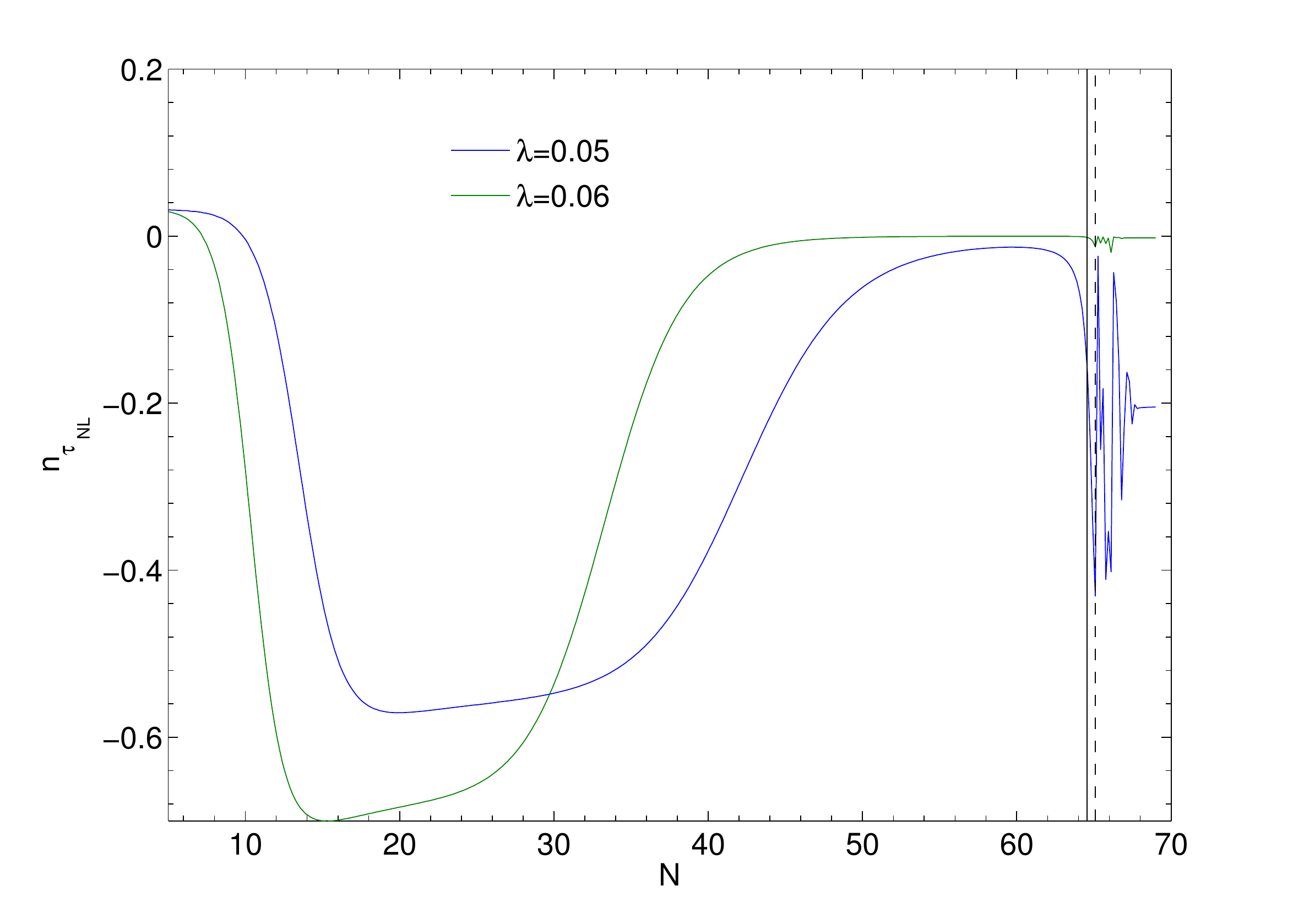}
	\end{tabular}	
	\caption{Potential: $W(\vp,\chi)=W_0\chi^2e^{-\lambda\vp^2/\Mp^2}$. \textit{Left panel}: The evolution of $\nfnl$.  \textit{Right panel}: The evolution of $\ntaunl$. The model parameters are $\lambda=\left\{0.05,\,0.06\right\}$, $\vps=10^{-3}\Mp$ and $\chis=16.0\Mp$, and $\Gchi=\sqrt{10^{-3}W_0}\Mp$. For $\lambda=0.05$, $\nfnl$ and $\ntaunl$ may be large enough to be observationally relevant, while for $\lambda=0.06$ the non--linear parameters are almost scale--independent.}
	\label{fig:runnings_quadexp}
\end{figure*}

Next we study the runnings of the non--linear parameters, $\nfnl$ and $\ntaunl$. In Fig.~\ref{fig:runnings_quadexp} we give the whole evolution of $\nfnl$ and $\ntaunl$ including reheating. The model parameters are $\lambda=\left\{0.05,\,0.06\right\}$, $\vps=10^{-3}\Mp$ and $\chis=16\Mp$, for the decay rate $\Gchi=\sqrt{10^{-3}W_0}\Mp$. For $\lambda=0.06$, we find that both $\nfnl$ and $\ntaunl$ are too small to be observationally relevant for CMB experiments, regardless of the decay rates $\Gchi$.  For $\lambda=0.05$, however, $\nfnl$ and $\ntaunl$ are much larger and of order $O(0.1)$, which could be potentially observed by Planck provided that the fiducial values of the non-linearity parameters are large enough. \\

\begin{figure*}[t]
	\begin{tabular}{cc}
		\includegraphics[width=0.48\linewidth]{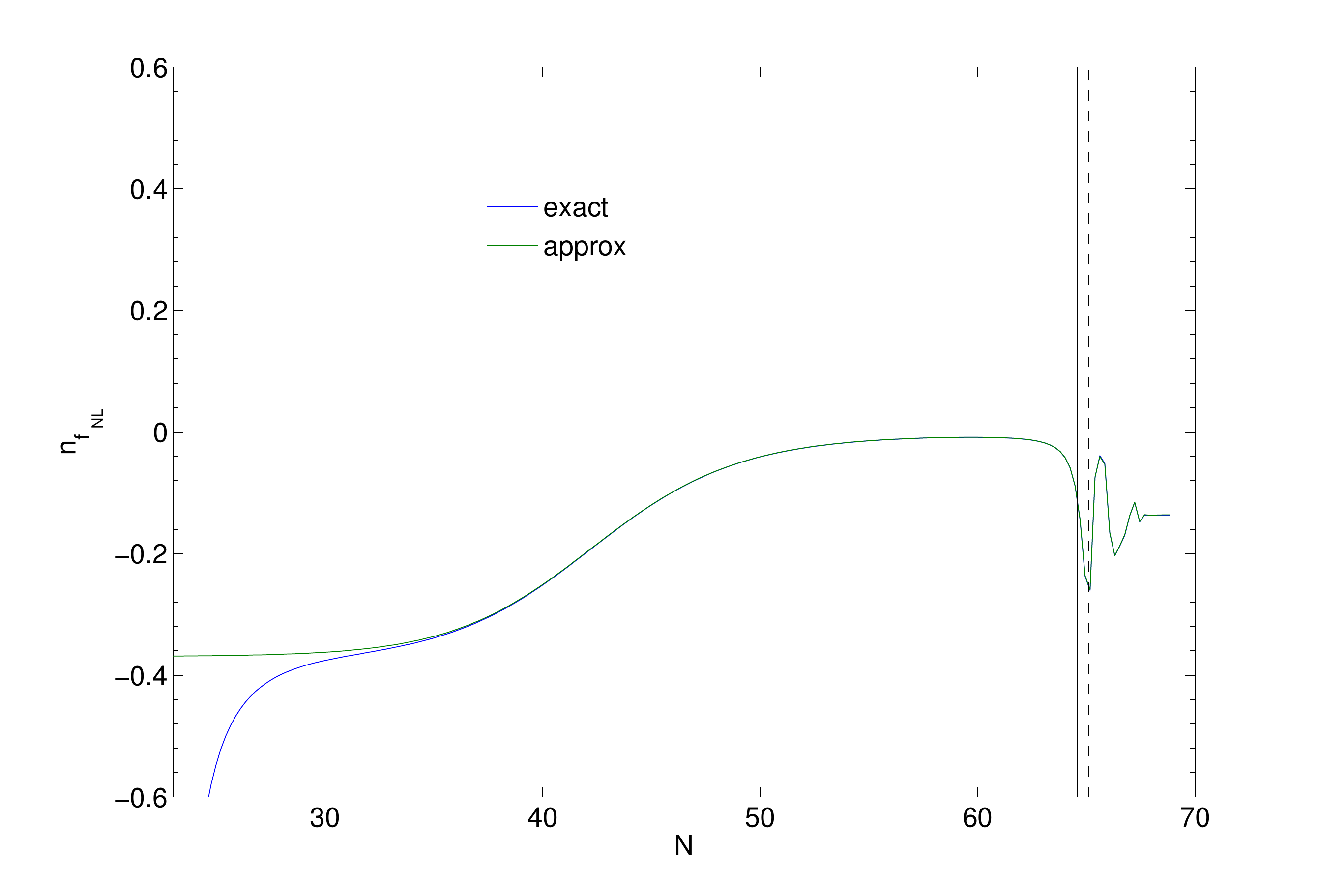}
		\includegraphics[width=0.45\linewidth]{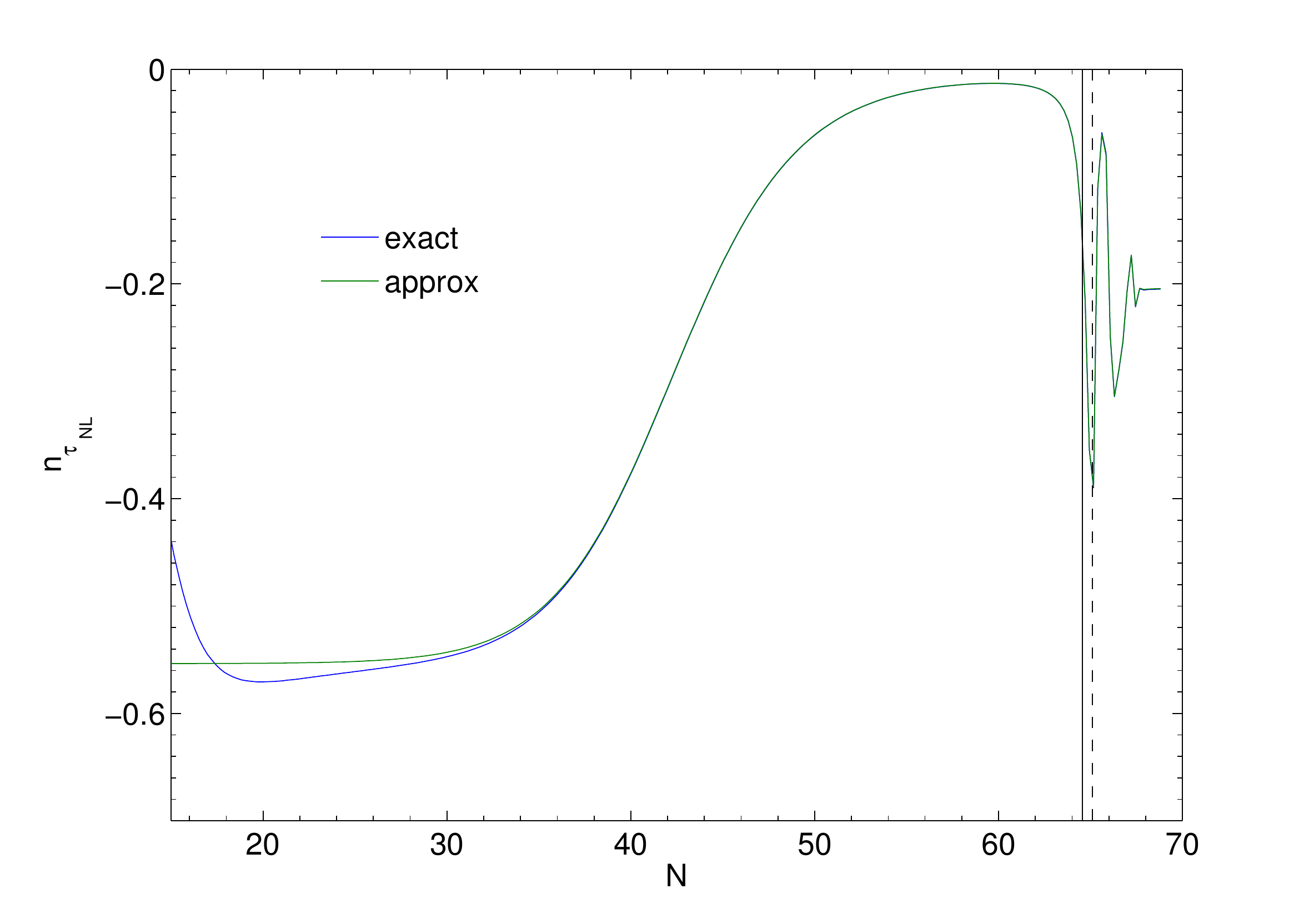}
	\end{tabular}	
	\caption{Potential: $W(\vp,\chi)=W_0\chi^2e^{-\lambda\vp^2/\Mp^2}$. Comparison of the exact Eqs.~(\ref{eq:nfnl_pot_2}-\ref{eq:ntaunl_pot_2}) and approximate formula Eq.~(\ref{eq:nfnl_ntaunl_approx}).  \textit{Left panel}: The evolution of $\nfnl$.  \textit{Right panel}: The evolution of $\ntaunl$. The model parameters are $\lambda=0.05$, $\vps=10^{-3}\Mp$ and $\chis=16.0\Mp$, for the decay rate $\Gchi=\sqrt{10^{-3}W_0}\Mp$. The two equations agree to a good approximation after about 30 e--folds of inflation.}
	\label{fig:runnings_quadexp_approx}
\end{figure*}

To understand why $\nfnl$ and $\ntaunl$ are much larger for $\lambda=0.05$, we first rewrite Eqs.~(\ref{eq:nfnl_pot}-\ref{eq:ntaunl_pot}) as
\bea
\nfnl = && -2(\nz - 1 +2\epsilon_*)  - \frac{5}{192}\frac{r^2}{\fnl} \nonumber \\ 
&& + \frac{5}{6\fnl}\left[\frac{4\eta_{IK*}N_{IJ}N_{J}N_{K} + \eta_{IJ*}N_{I}N_{J} + (W_{IJK}/W)_*N_{I}N_{J}N_{K}}{(N_{L}N_{L})^2}\right]\,, 
\label{eq:nfnl_pot_2} \\
 \ntaunl = && -3(\nz - 1 +2\epsilon_*) - \frac{1}{256}\frac{r^{3}}{\taunl} \nonumber \\ 
&& + \frac{2}{\taunl}\left[\frac{2\eta_{JL*}N_{IJ}N_{IK}N_{L}N_{K} + \eta_{IJ*}N_{I}N_{J} + \eta_{IJ*}N_{JL}N_{IK}N_{L}N_{K}+(W_{IJL}/W)_*N_{IK}N_{J}N_{K}N_{L}}{(N_{M}N_{M})^3}\right] \,,
\label{eq:ntaunl_pot_2}
\eea
using $r=\frac{8}{N_{I}N_{I}}$ where $r$ is the tensor--to--scalar ratio~\cite{Vernizzi2006NonGaussianities}. From this, it is not difficult to see that the second terms in the first line of both equations are small in general because of the tight observational constraint imposed on r, namely $r<0.38$ at $95\%\,CL$~\cite{Bennett:2012fp}. Making use of the approximate formulae for $\fnl$ and $\taunl$, the dominant terms in Eqs.~(\ref{eq:nfnl_pot_2}-\ref{eq:ntaunl_pot_2}) are
\bea \ntaunl\simeq\frac32\nfnl\simeq -3(\nz - 1 + 2\epsilon_*) + 6\eta^{*}_{\vp\vp} \simeq 6\eta^{*}_{\vp\vp}\left(\frac{\Nchi^2}{\Nvp^2+\Nchi^2}\right)\,, \label{eq:nfnl_ntaunl_approx} \eea
where we have assumed slow--roll at horizon--crossing such that $\left(\frac{W_{IJK}}{W}\right)_*\ll O(1)$ and used Eq.(43) in~\cite{Leung:2012ve}, i.e. $\nz - 1 + 2\epsilon_* \approx -2\eta^{*}_{\vp\vp}\left(\frac{\Nvp^2}{\Nvp^2 + \Nchi^2}\right)$. We have also assumed that the numerators in the square brackets in Eqs.~(\ref{eq:nfnl_pot_2}-\ref{eq:ntaunl_pot_2}) are dominated by the $\Nvp$ and $\Nvpvp$ terms. In Fig.~\ref{fig:runnings_quadexp_approx}, we show the comparison between the exact Eqs.~(\ref{eq:nfnl_pot_2}-\ref{eq:ntaunl_pot_2}) and the approximate formula Eq.~(\ref{eq:nfnl_ntaunl_approx}). From this, we can see the approximate formula agrees very well with the full expressions after about $30$ e-folds of inflation, even during the reheating phase.
From Eq.~(\ref{eq:nfnl_ntaunl_approx}), one may see that the runnings are relatively large when $\Nvp\sim\Nchi$, which is the case when $\lambda=0.05$, but very small when $|\Nvp|\gg|\Nchi|$, which is the case when $\lambda=0.06$. Notice that if $|\Nvp|\gg|\Nchi|$, the runnings are driven to zero and hence become independent of the decay rate.

\begin{figure}
\begin{center}
       	\includegraphics[width=0.45\linewidth]{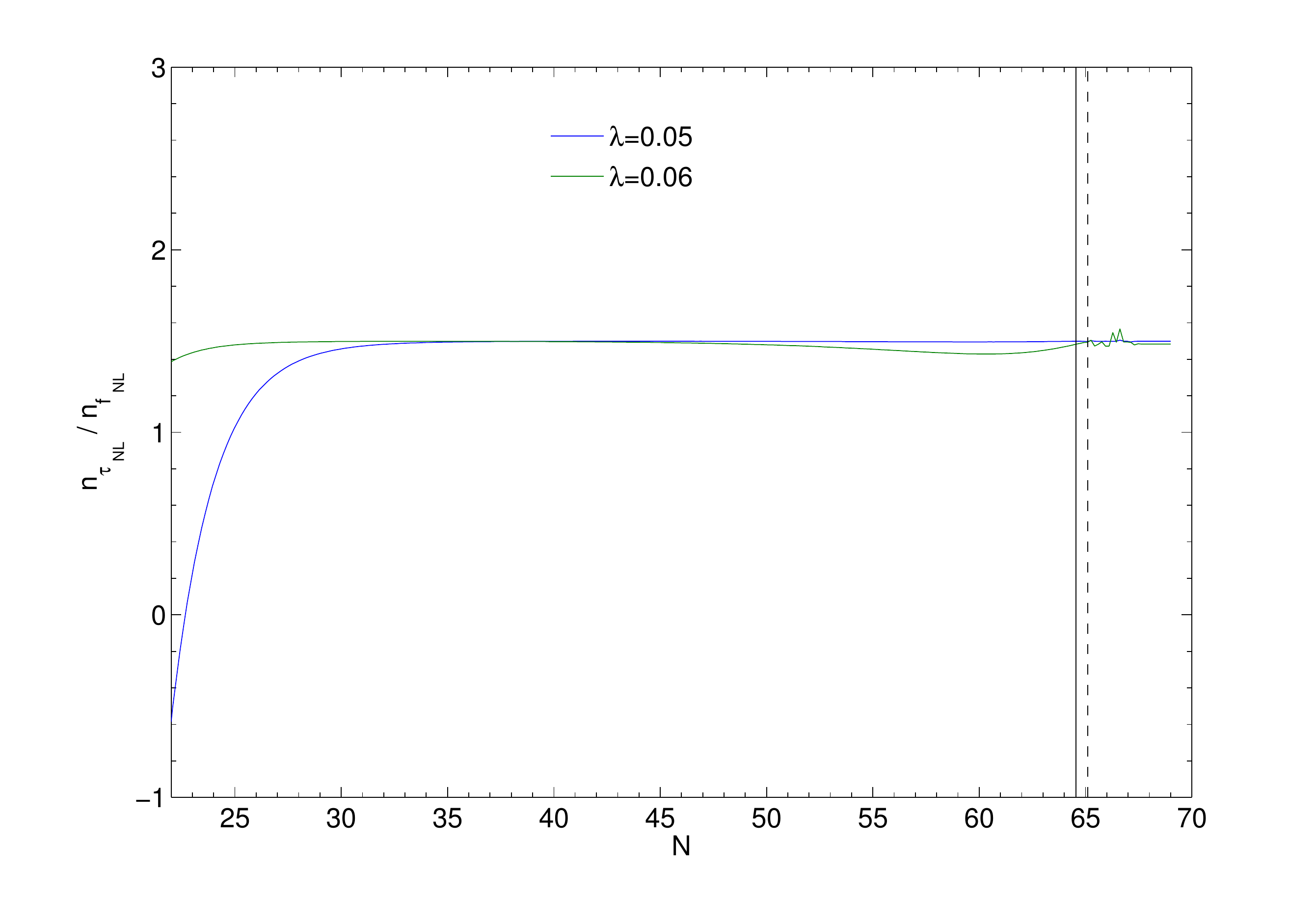} \nonumber 
      	\caption{Potential: $W(\chi,\vp) = W_{0}\chi^{2}e^{-\lambda\vp^2/\Mp^2}$. The evolution of the ratio $\ntaunl/\nfnl$ until the completion of reheating. The model parameters are $\lambda=\left\{0.05,\,0.06\right\}$, $\vps=10^{-3}Mp$, $\chis=16\Mp$ and $\Gchi=\sqrt{0.3W_0}\Mp$. The ratio settles to $3/2$ quickly after about 30 e--folds of inflation from Hubble exit, satisfying the consistency relation Eq.~(\ref{eq:running_2field_local}). } 
	\label{fig:nfnl_ntaunl_quadexp}
\end{center}
\end{figure}

Whether $\nfnl$ and $\ntaunl$ are of a detectable level or not, we find that they satisfy the consistency relation
\bea
n_{\taunl} = \frac{3}{2} n_{\fnl} \,,
\label{eq:running_2field_local}
\eea 
regardless of $\Gchi$ and thus the reheating timescale. This relation was first observed to hold for some classes of two field models in~\cite{Byrnes:2010ft}. We provide an example of this scaling behaviour in Fig.~\ref{fig:nfnl_ntaunl_quadexp}, where we observe it to hold throughout most of the evolution, except partly during the first $20$ e-folds (of course this evolution is not itself observable, only the final values). We will discuss this relation in further details in Section~\ref{sec:disc}. \\
%
%-------------------------------------------------------------------------------------------------------------
\linebreak
In Table~\ref{tab:oneminstats} we summarise the results, showing the comparison between the primordial observables evaluated at the end of inflation and at the end of reheating. This is one of the main results of this paper. Notice the different qualitative behaviour for the non--linear parameters in the models for different $\lambda$, where the magnitudes of $\fnl$ and $\taunl$ decrease with larger $\Gchi$ for $\lambda=0.05$, but increase for $\lambda=0.06$.  In general, the final values of the non--linear parameters at the completion of reheating is different from the end of inflation values, whilst $\gnl$ remains small $\ll O(100)$ in this model which is unlikely be observed in future experiments. The runnings $\nfnl$ and $\ntaunl$ are large in the case $\lambda=0.05$ and are redder for larger $\Gchi$.

%-------------------------------------------------------------------------------------------------------------
\renewcommand*\arraystretch{1.2}
%-------------------------------------------------------------------------------------------------------------
\begin{table}[h!]
\vspace{5pt}
    \begin{center}
        \subtable{
\begin{tabular}{c|c|c|c|c|c}
\multicolumn{6}{c}{End of Inflation, $\lambda=0.05$} \\
\hline
\hline
   $-$  &  $\fnl$ &  $\taunl$  &    $\gnl$   &  $\nfnl$  & $\ntaunl$ \\
\hline
 $-$    &    $-34.1$   &	  $2.34\times 10^3$   &       $-49.6$	    &        $-0.105$  & $-0.158$  \\
\hline
\multicolumn{6}{c}{End of Reheating,$\lambda=0.05$} \\
   $\Gchi$  &    $\fnl$ &  $\taunl$  &    $\gnl$   &  $\nfnl$  & $\ntaunl$ \\
\hline
 $\sqrt{10^{-5}}$    &    $-33.4$   &	  $2.25\times 10^3$   &       $-13$	    &        $-0.105$  & $-0.157$  \\
 $\sqrt{10^{-3}}$    &    $-31.5$   &	  $2.27\times 10^3$  &       $-11.6$	    &        $-0.137$   &  $-0.205$ \\
 $\sqrt{10^{-1}}$    &    $-26.9$   &         $2.01\times 10^3$   &       $-9.96$	    &        $-0.177$   &  $-0.266$  \\
                \end{tabular}\centering
                }
\hspace{10mm}
	\subtable{
\begin{tabular}{c|c|c|c|c|c}
\multicolumn{6}{c}{End of Inflation, $\lambda=0.06$} \\
\hline
\hline
   $-$  &  $\fnl$ &  $\taunl$  &    $\gnl$   &  $\nfnl$  & $\ntaunl$ \\
\hline
 $-$    &    $-5.93$   &   $50.7$   &       $9.86$            &        $-1.0\times 10^{-3}$  & $-1.5\times 10^{-3}$  \\
\hline
\multicolumn{6}{c}{End of Reheating, $\lambda=0.06$} \\
   $\Gchi$  &    $\fnl$ &  $\taunl$  &    $\gnl$   &  $\nfnl$  & $\ntaunl$ \\
\hline
 $\sqrt{10^{-5}}$    &    $-4.35$   &      $28.1$   &       $-2.41$           &        $-9.1\times10^{-4}$  & $-1.3\times10^{-3}$  \\
 $\sqrt{10^{-3}}$    &    $-5.54$   &      $44.5$  &       $-2.62$            &        $-1.4\times 10^{-3}$   &  $-2.1\times10^{-3}$ \\
 $\sqrt{10^{-1}}$    &    $-7.14$   &   $73.9$   &       $-2.96$     &        $-2.3\times10^{-3}$   &  $-3.4\times10^{-3}$  \\
                \end{tabular}               
                }
\caption{Statistics of $\zeta$ for $W(\vp,\chi)=W_0\chi^2e^{-\lambda\vp^2/\Mp^2}$ for different decay rates. All decay rates are in units of $\sqrt{W_0}\Mp$. We give values computed at the end of inflation ($N_e$) and at the completion of reheating (final) where $\zeta$ becomes conserved. The model parameters are $\lambda=0.05$ \textit{(Left panel)} and $0.06$ \textit{(Right panel)}, $\vps=10^{-3}\Mp$ and $\chis=16.0\Mp$. }\label{tab:oneminstats}
    \end{center}
\end{table}
%

%-------------------------------------------------------------------------------------------------------------
\subsection{Two Minima}\label{sec:twomins}

Next we consider a model where the potential has minima in both field directions. Both fields can now decay to the effective radiation fluid and be directly involved in reheating. The model considered is the effective two--field description of axion N--flation, where the potential given by~\cite{elliston:2011}
\be
\label{eq:effectiveNflation}
W(\vp,\chi)=W_0\left[\frac{1}{2}m^2\chi^2 + \Lambda^4\left(1 -{\rm cos}\left(\frac{2\pi}{f}\vp\right)\right)\right]\,.
\ee
The axion $\vp$, is described by its decay constant $f$ and its potential energy scale $\Lambda^4$. To generate a large non--Gaussianity, we must have $\vp$ close to the ``hilltop'' at horizon--crossing~\cite{elliston:2011}. In this configuration, the second field $\chi$, drives inflation.\footnote{We have also studied the models $W=W_0\left[\frac{\lambda}{4!}\chi^4 + \Lambda^4\left(1 -{\rm cos}\left(\frac{2\pi}{f}\vp\right)\right)\right]$ and $ W=W_0\left(\frac{1}{4!}g\chi^4 + V_0 + h\vp+ \frac{1}{3!}\lambda\vp^3 +\frac{1}{4!}\mu\vp^4 \right)$. Similar conclusions were found in these models.}

%-------------------------------------------------------------------------------------------------------------
\subsubsection{Trispectrum} \label{sec:trispectrum_quadtwomins}

\begin{figure}
        \begin{tabular}{cc}
		\includegraphics[width=0.45\linewidth]{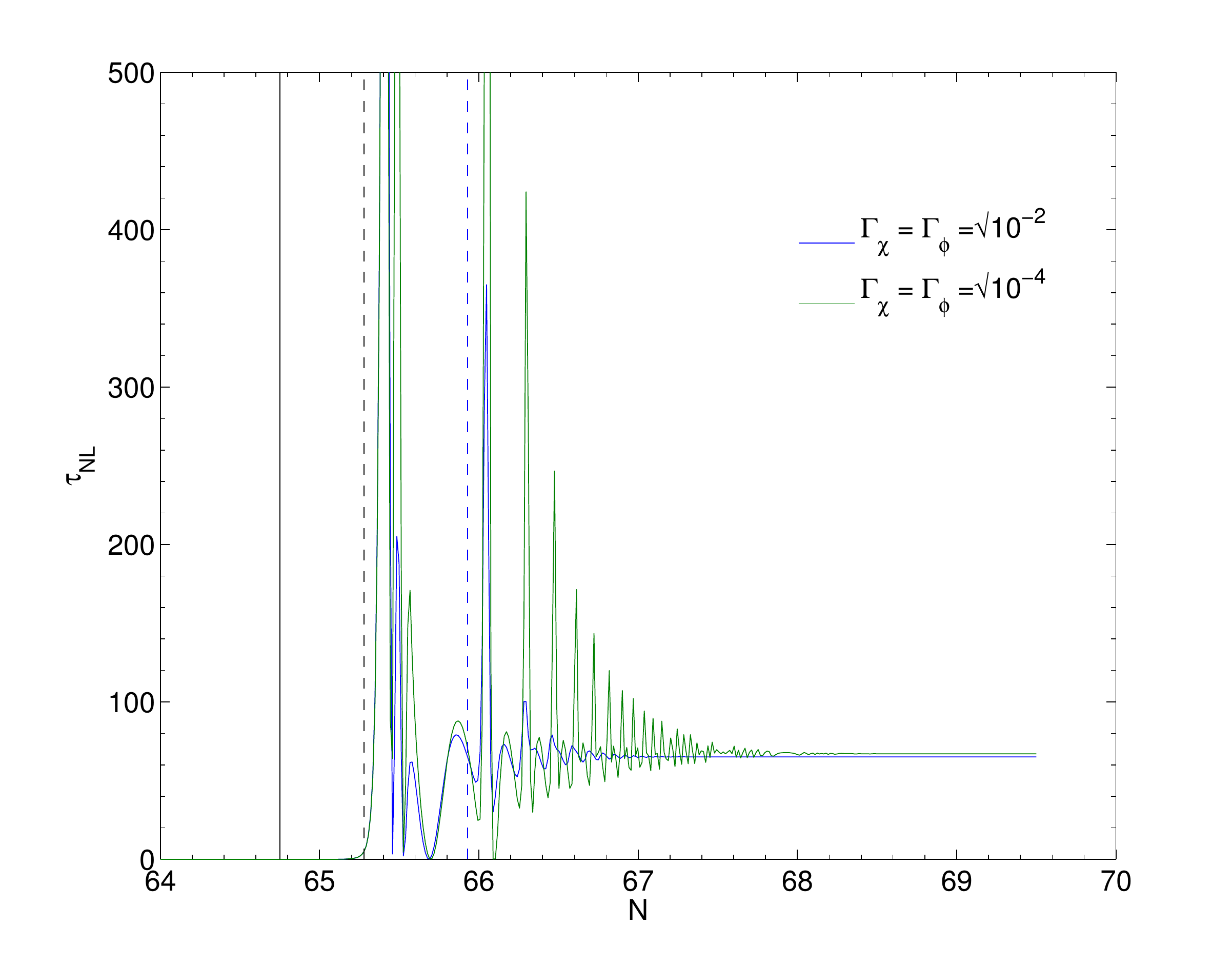}
		\includegraphics[width=0.45\linewidth]{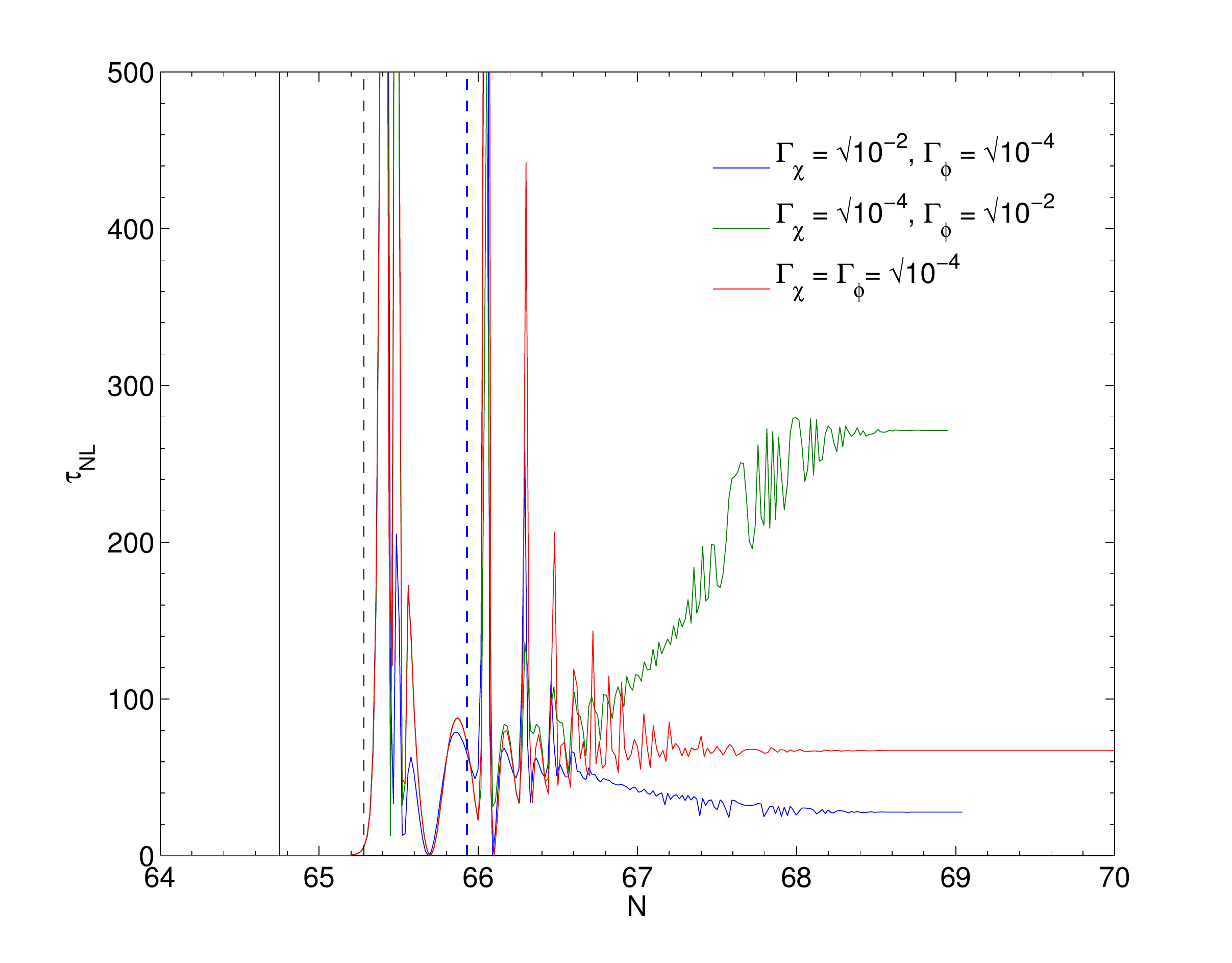}
	\end{tabular}
      	\caption{Potential: $W (\vp,\chi) = W_0 [\frac{1}{2}m^2\chi^2 + \Lambda^4(1-{\rm cos}(\frac{2\pi}{f}\vp))]$. The evolution of $\taunl$ during post--inflationary period. The model parameters are $\Lambda^4=m^2f^2/4\pi^2$, $\vps = (\frac{f}{2\Mp}-0.001)\Mp, \chis = 16\Mp$ and $f=m=\Mp$. For these model parameters, the $\chi$ field minimises before the $\vp$ field.  \textit{Left panel}: Equal decay rates, $\Gchi=\Gvp$. \textit{Right panel}: Unequal decay rates, $\Gchi\neq\Gvp$. The solid vertical line denotes the end of inflation, $N_{\rm e}$, and the dashed lines denote the start of reheating, $N|_{\rm\vp=0}$ (blue) and $N|_{\rm\chi=0}$ (black), respectively, in this figure and all subsequent figures for this two minima model.  Notice that $\taunl$ changes by two orders of magnitude during reheating. Also, $\taunl$ is sensitive to $\Gchi$ and $\Gvp$ if there is a hierarchy between the two decay rates.}
	\label{fig:tauNL_axion}
\end{figure}

The model parameters we consider are $\Lambda^4=m^{2}f^{2}/4\pi^2$, $\vps = (\frac{f}{2\Mp}-0.001)\Mp, \chis = 16\Mp$ and $f=m=\Mp$. Each of $\fnl$, $\taunl$ and $\gnl$ are negligible during inflation as the axion $\vp$ is sufficiently light that it remains almost frozen near the top of the ridge. In this sense, this scenario is similar to the curvaton model. Things are different after inflation ends however. When inflation ends, the axion $\vp$ starts rolling down the ridge, producing a negative spike in $\fnl$. $\fnl$ then evolves to positive value when the $\vp$ field converges to its minimum~\cite{Leung:2012ve}. It is similar for $\taunl$, except $\taunl$ is always positive. In Fig.~\ref{fig:tauNL_axion} we give the evolution of $\taunl$ during reheating for various combinations of $\Gchi$ and $\Gvp$. \\ 

Similar to $\fnl$ as studied in~\cite{Leung:2012ve}, we find that although the final value of $\taunl$ is different from that at the end of inflation, it is almost completely insensitive to the decay rates if $\Gchi=\Gvp$. Things are different however if there is a mild hierachy between $\Gchi$ and $\Gvp$. When $\Gchi\neq\Gvp$, the final value of $\taunl$ does depend on the reheating timescale. Compared to the value where $\Gchi=\Gvp$, it grows for $\Gvp>\Gchi$ and decays for $\Gchi>\Gvp$.   

\begin{figure*}[!htb]
	\begin{tabular}{cc}
		\includegraphics[width=0.50\linewidth]{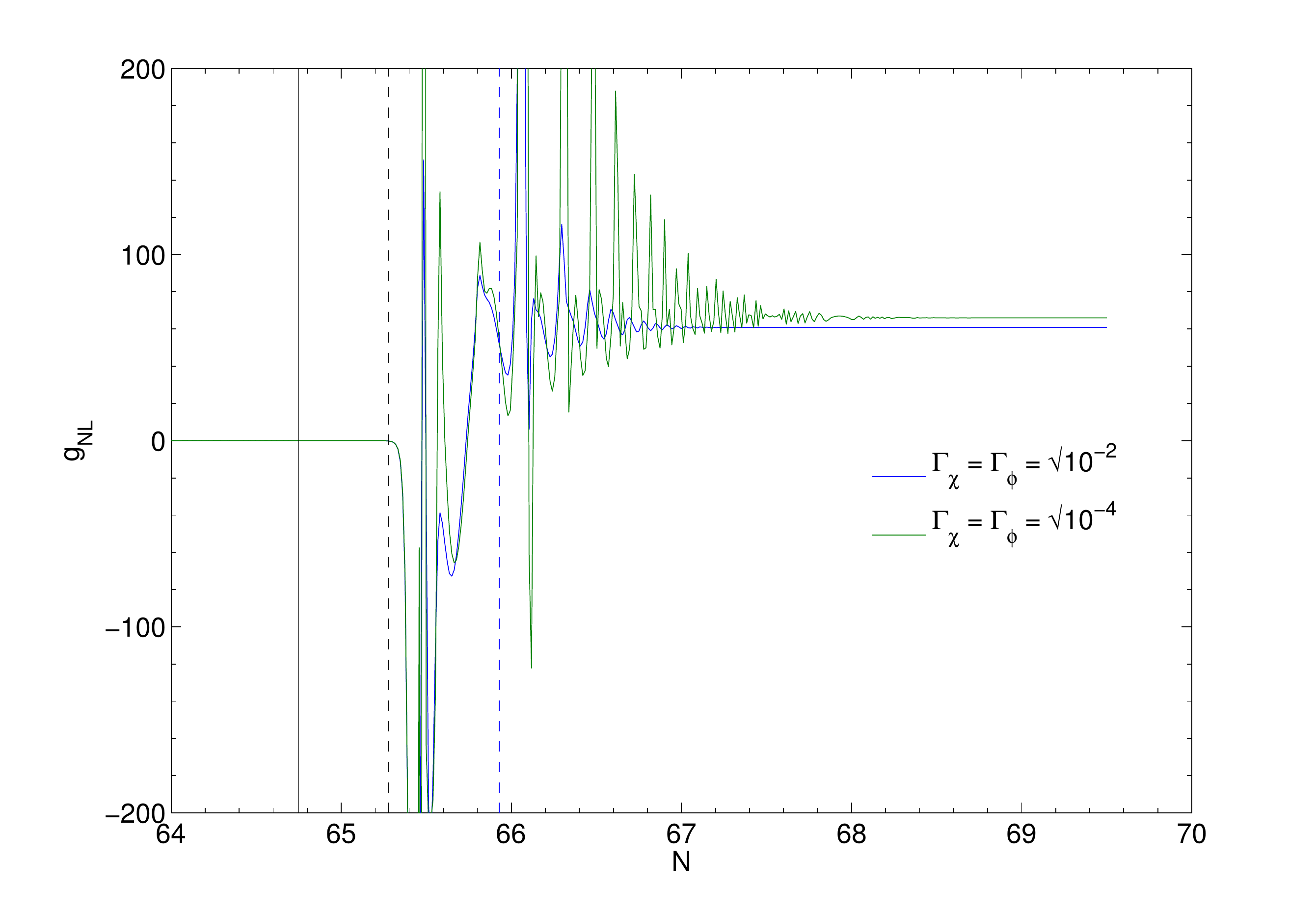}
		\includegraphics[width=0.50\linewidth]{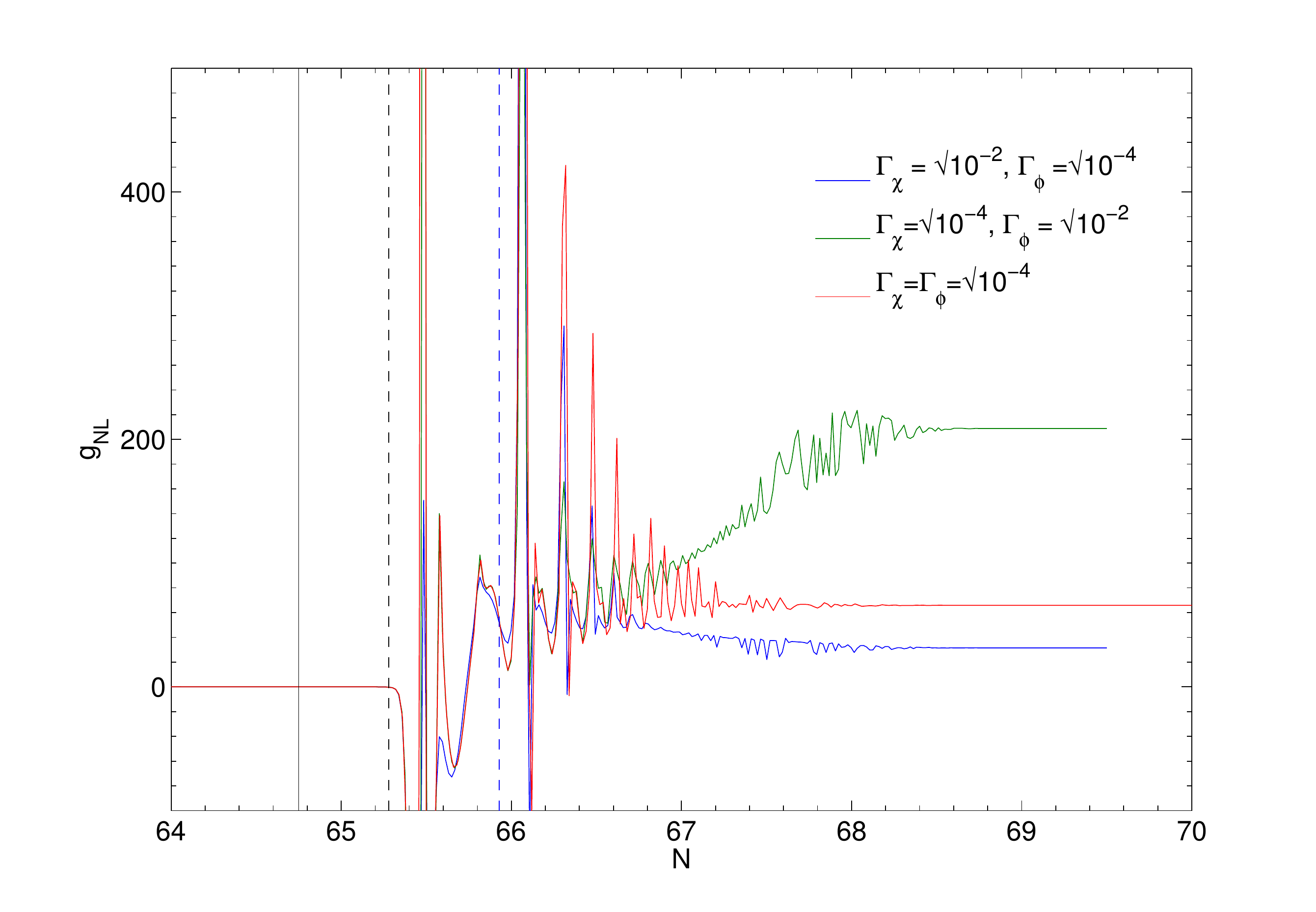}
	\end{tabular}	
        \caption{Potential: $W(\chi,\vp) = W_0 \left\{\frac{1}{2}m^2\chi^2 + \Lambda^4\left[1-{\rm cos}(\frac{2\pi}{f}\vp)\right]\right\}$. The model parameters are $\Lambda^4=m^2f^2/4\pi^2$, $\vps = (\frac{f}{2\Mp}-0.001)\Mp, \chis = 16\Mp$ and $f=m=\Mp$. \textit{Left panel}: Equal decay rates, $\Gchi=\Gvp$; \textit{Right panel}: Unequal decay rates, $\Gchi\neq\Gvp$. Similar to $\taunl$, $\gnl$ changes by two orders of magnitude during reheating and is more sensitive to the decay rates whenever there is a hierarchy between them.} 
	\label{fig:gNL_axion}
\end{figure*}

\begin{figure*}[!htb]
	\begin{tabular}{cc}
		\includegraphics[width=0.50\linewidth]{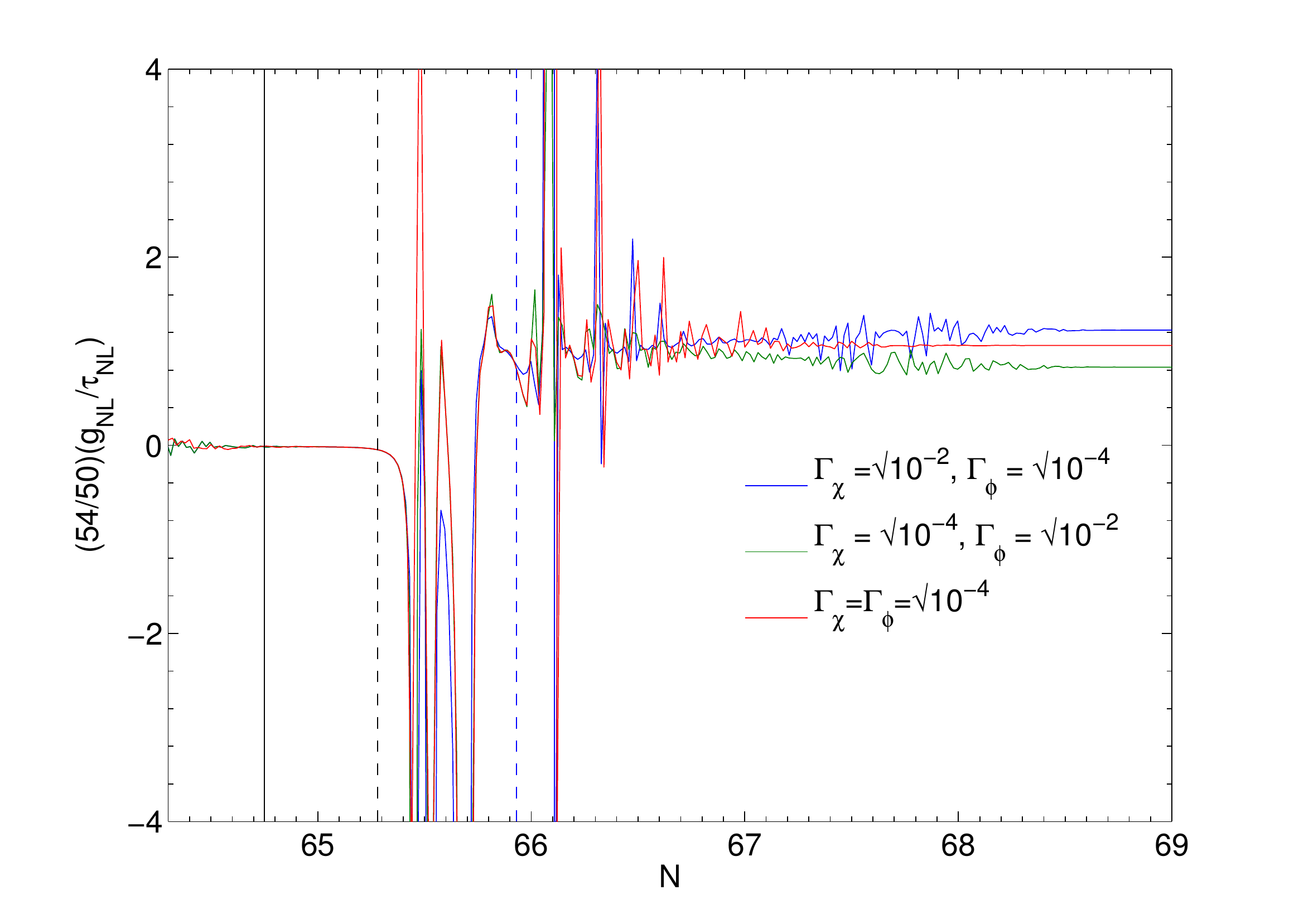}
		\includegraphics[width=0.50\linewidth]{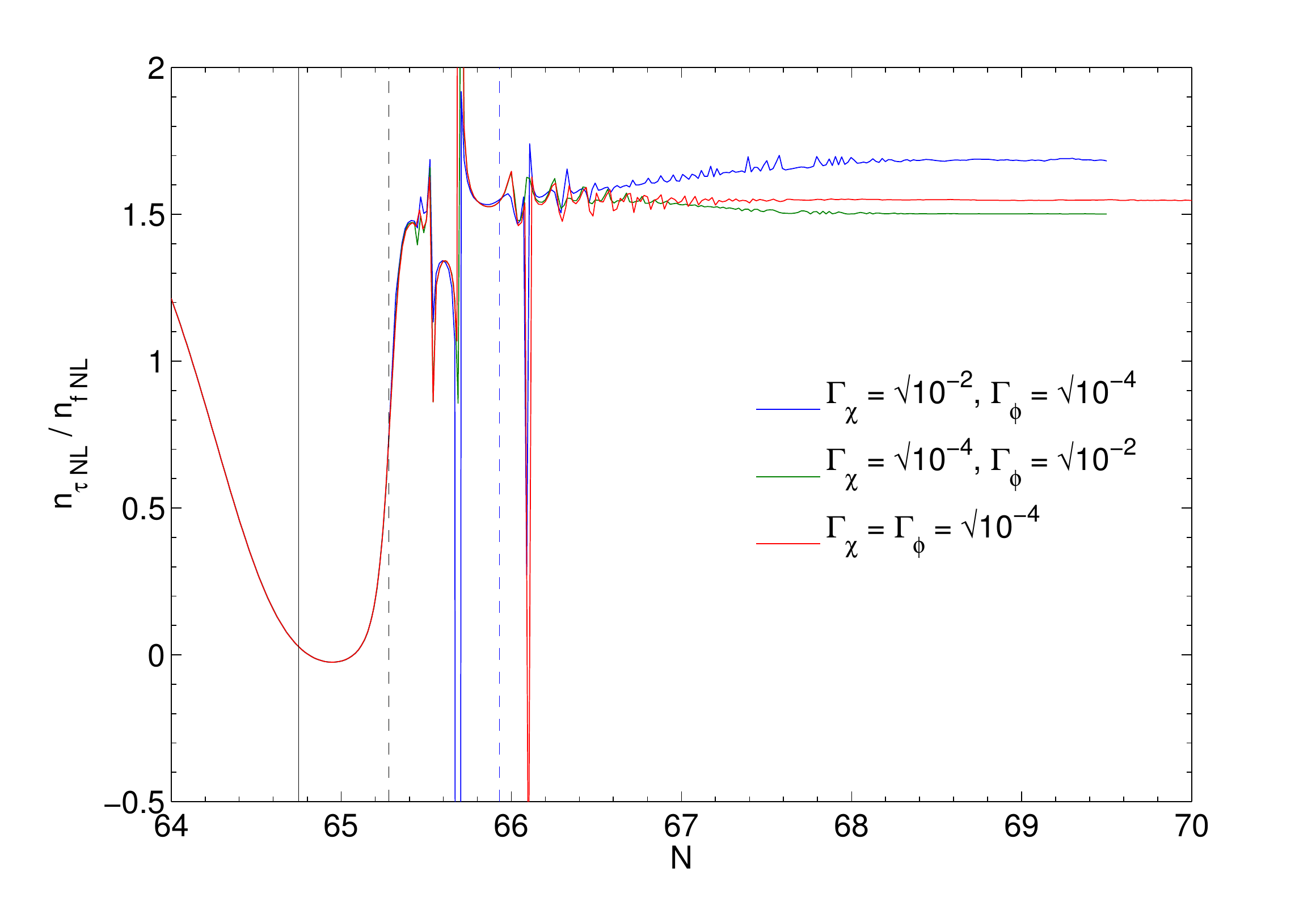}
	\end{tabular}	
        \caption{Potential: $W(\chi,\vp) = W_0 \left\{\frac{1}{2}m^2\chi^2 + \Lambda^4\left[1-{\rm cos}(\frac{2\pi}{f}\vp)\right]\right\}$. The model parameters are $\Lambda^4=m^2f^2/4\pi^2$, $\vps = (\frac{f}{2\Mp}-0.001)\Mp, \chis = 16\Mp$ and $f=m=\Mp$. \textit{Left panel}: The evolution of the ratio $(27/25)(\gnl/\taunl)$ during reheating for different combinations of decay rates. \textit{Right panel}: evolution of the ratio $\ntaunl/\nfnl$ during the post--inflationary period. Notice that the relations Eqs.~(\ref{eq:running_2field_local}) and (\ref{eq:gNL_tauNL_relation_sum}) are satisfied only after inflation ends. Both relations are only mildly violated when $\Gchi\neq\Gvp$}
	\label{fig:gNL_tauNL_axion}
\end{figure*}

Unlike the previous model in Section~\ref{sec:onemin}, no scaling relation is found between $\Nvpvp$ and $\Nvp$. Yet we can still make use of the observations that $\Nvpvp$ and $\Nvp$ dominate over $\Nchichi, \Nchivp$ and $\Nchi$ respectively to rewrite Eq.~(\ref{eq:taunl_Dn}) as
\bea
\taunl \approx \frac{\Nvpvp^2}{\Nvp^4} \label{eq:tauNL_approx_axion} \,.
\eea
For $\gnl$, things are similar to $\fnl$ and $\taunl$. In Fig.~\ref{fig:gNL_axion} we give the evolution of $\gnl$ for different combinations of $\Gchi$ and $\Gvp$, with the same model parameters. While the final values of $\gnl$ at the end of reheating are different from that at the end of inflation, they are almost completely insensitive to $\Gchi$ and $\Gvp$ unless there is a mild hierachy between the decay rates.
Again with a hierachy between the third order $\Dn$ derivatives found, $\gnl$ can be well approximated by
\bea
\gnl \approx \frac{25}{54}\frac{\Nvpvpvp}{\Nvp^3} \label{eq:tauNL_approx_axion} \,.
\eea
During slow--roll, Elliston et.al.~\cite{Elliston2012Large} have shown that $\gnl$ is roughly of the same order as $\taunl$ and the following relation holds
\bea
\frac{27}{25}g_{NL} \approx \taunl  \label{eq:gNL_tauNL_relation_sum} \,,
\eea
for non-vacuum dominated sum--separable potentials, given that $\taunl$ is large. Here we find that this holds beyond the slow--roll regime and during reheating for a range of mass ratios between the axion and inflaton where they both minimise after the end of inflation. For example, from Fig.\ref{fig:gNL_tauNL_axion}, we can see that this relationship is only mildly violated when $\Gchi\neq\Gvp$. 

\subsubsection{Runnings of non--linear parameters, $\nfnl$ and $\ntaunl$}

We now turn our attention to the study of the runnings $\nfnl$ and $\ntaunl$ in this model. Similar results are found as in the one minimum case where $\lambda=0.06$, i.e. both runnings are small and $\frac{3}{2}\nfnl\simeq\ntaunl$, except the relation Eq.~(\ref{eq:running_2field_local}) may be mildly violated when $\Gchi\gg\Gvp$. Yet one should notice that the relation Eq.~(\ref{eq:running_2field_local}) does not hold throughout the entire evolution, but only after inflation ends when both fields start oscillating. Therefore one would end up in a completely different conclusion that Eq.~(\ref{eq:running_2field_local}) does not hold for the model if $\nfnl$ and $\ntaunl$ are evaluated only up to inflation ends.\\

In Table~\ref{tab:twominstats} we summarise these results, showing the comparison between the primordial observables evaluated at the end of inflation and at the end of reheating. This is one of the main results of this paper, which clearly show the non--linear parameters in multifield models strongly depend on the reheating timescale in general. Notice the large differences between the statistics evaluated at the end of inflation, compared to the end of reheating. This is because the axion field only begins to roll after inflation has ended and so until this point, the observables do not evolve appreciably. 

%-------------------------------------------------------------------------------------------------------------
\renewcommand*\arraystretch{1.2}
%-------------------------------------------------------------------------------------------------------------
\begin{table}[h!]
\vspace{5pt}
    \begin{center}
\begin{tabular}{c|c|c|c|c|c|c}
\multicolumn{7}{c}{End of Inflation} \\
\hline
\hline
   $-$  &  $-$   & $\fnl$ &  $\taunl$  &    $\gnl$   &  $\nfnl$  & $\ntaunl$ \\
\hline
 $-$    &  $-$  &   $5.9\times 10^{-3}$   &	  $1.3\times 10^{-3}$   &       $-3.1\times 10^{-5}$	    &        $1.7\times 10^{-2}$  & $4.9\times 10^{-4}$  \\
\multicolumn{7}{c}{End of Reheating}  \\
\hline
\hline
  $\Gchi$   &  $\Gvp$ &  $\fnl$  & $\taunl$  &    $\gnl$   &  $\nfnl$   &  $\ntaunl$\\
\hline
 	$0$  &  	$0$       &       	$6.88$  &  	$0.69\times 10^2$  &  	$0.63\times 10^2$  &  	$-1.2\times 10^{-6}$  &  	$-1.8\times 10^{-6}$    \\
 	$\sqrt{10^{-2}}$  &  	$\sqrt{10^{-2}}$       &       	$6.59$  &  	$0.76\times 10^{2}$  &  	$0.55 \times 10^2$  &  	$-9.3\times 10^{-7}$  &  	$-1.6\times 10^{-6}$    \\
 	$\sqrt{10^{-2}}$  &  	$\sqrt{10^{-4}}$       &       	$4.37$  &  	$0.29\times 10^2$  &  	$0.29\times 10^2$  &  	$-7.2\times 10^{-7}$  &  	$-1.2\times 10^{-6}$    \\
 	$\sqrt{10^{-4}}$  &  	$\sqrt{10^{-2}}$       &       	$13.66$  &  	$2.75\times 10^{2}$  &  	$1.91\times 10^{2}$  &  	$-2.5\times 10^{-6}$  &  	$-3.7\times 10^{-6}$    \\
     $\sqrt{10^{-4}}$  &  	$\sqrt{10^{-4}}$       &       	$6.83$  &  	$0.68\times 10^2$  &  	$0.59\times 10^2$  &  	$-1.1\times 10^{-6}$  &  	$-1.7\times 10^{-6}$  \\
                \end{tabular}\centering
\caption{Statistics of $\zeta$ for $W(\vp,\chi)=W_0\left[\frac{1}{2}m^2\chi^2 + \Lambda^4\left(1 -{\rm cos}\left(\frac{2\pi}{f}\vp\right)\right)\right]$ for different decay rates. All decay rates are in units of $\sqrt{W_0}\Mp$. We give values computed at the end of inflation ($N_e$) and at the completion of reheating (final) where $\zeta$ is conserved. The model parameters are $\Lambda^4=m^2f^2/4\pi^2$, $\vps = (\frac{f}{2\Mp}-0.001)\Mp, \chis = 16\Mp$ and $f=m=\Mp$. Note that the values in the row where $\Gchi=\Gvp=0$ do not correspond to end of reheating since the decay rates are zero. However an adiabatic limit is still reached as both $\vp$ and $\chi$ behave as matter fluids when oscillating about their minima.}
\label{tab:twominstats}
    \end{center}
\end{table}

%-------------------------------------------------------------------------------------------------------------
\section{Discussion}\label{sec:disc}

\subsection{Relation between $\taunl$ and $\fnl$}

In general, $\gnl$, $\taunl$ and $\fnl$ are functions of external momenta $\mathbf{k_i}$ which cannot be compared directly. Yet in canonical models, when the non--gaussianity is large, it is dominated by the shape independent parts. It is thus reasonable to compare the non--linear parameters directly in such models.

The Suyama-Yamaguchi inequality~\cite{Suyama2008NonGaussianity}, for instance, relates $\fnl$ in the squeezed limit $(\mathbf{k_1}\rightarrow 0)$ to $\taunl$ in the collapsed limit $(\mathbf{k_1+k_2}\rightarrow 0)$
\bea
\taunl\geq(\frac{6}{5}\fnl)^2 \label{eq:SYineq}\,.
\eea
This inequality has been studied and verified extensively in the literature, see e.g.~\cite{Suyama:2010uj,Lewis:2011au,Smith:2011if,Sugiyama:2012tr,Assassi:2012zq,Kehagias:2012pd,Tasinato:2012js}. For a recent review, see~\cite{Rodriguez:2013cj}. While the equality in Eq.~(\ref{eq:SYineq}) holds for single-source models, multifield models in general give $\taunl>(6\fnl/5)^2$~\cite{Suyama:2010uj}.

Recently, Peterson et.al.\cite{Peterson:2010mv} have shown that $\taunl$ is not much larger than $\fnl^2$ in two--field canonical models in general by applying both the slow--roll and slow-turn approximations, except in cases of excessive fine-tuning. It was later verified by Elliston et.al.~\cite{Elliston2012Large} to hold also for separable potentials. 

In canonical two--field models, a large non--Gaussianity is typically generated by having one of the fields rolling down an extreme point like a ridge or a valley. During slow--roll, $\fnl$ and $\taunl$ can be approximated by
\bea
\fnl\approx \frac{6}{5}\frac{\Nvpvp\Nvp^2}{(\Nvp^2+\Nchi^2)^2}\,, \label{eq:fnl_2field} \\
\taunl\approx \frac{\Nvpvp^2\Nvp^2}{(\Nvp^2+\Nchi^2)^3}\,.   \label{eq:taunl_2field}
\eea
From Eqs.~(\ref{eq:fnl_2field}-\ref{eq:taunl_2field}), we then have
\bea
\frac{\taunl}{(6/5)^2\fnl^2} \approx 1 + \frac{\Nchi^2}{\Nvp^2}\,. \label{eq:taunl_fnl_approx}
\eea
As a result, in order to have $\taunl\gg\fnl^2$ and $|\fnl|>O(1)$, one typically needs $|\Nvp|\ll|\Nchi|$ while $|\Nvpvp|\gg|\Nchichi|,|\Nchivp|$. This is highly non--trivial for any function of $N$, and in general is difficult to accommodate in canonical two--field models.

Yet field dynamics during reheating is very different from slow--roll inflation and thus one might expect reheating would significantly change this conclusion. First, it is not obvious that the same approximation Eq.~(\ref{eq:fnl_2field}-\ref{eq:taunl_2field}) would hold after reheating. Even if the approximation holds, it is possible that $N$ develop additional non--trivial dependence on $\vps$ and $\chis$ such that $\taunl$ is greatly enhanced during reheating compared to $\fnl^2$. However, in the models we study, we found that the conclusion that $\taunl$ is not much larger than $\fnl^2$ seems to hold after reheating for a large range of decay rates $\Gvp$ and $\Gchi$, as shown in Table~~\ref{tab:oneminstats} and \ref{tab:twominstats}.

\subsection{Relation between $\taunl$ and $\gnl$}

For non--vacuum dominated sum--separable potentials, by making use of the analytic expressions for $\Dn$ derivatives, Elliston et.al.~\cite{Elliston2012Large} have shown that $\gnl$ and $\taunl$ are of the same order 
\bea
\frac{27}{25}\gnl \approx \taunl \label{eq:nonvacuum_sumsep_gNL_tauNL}\,,
\eea
in the absence of significant terms beyond quadratic order in the potential. The effective N--flation model we studied in Section~\ref{sec:twomins} is of this type. We found that given $\taunl$ is large, this relation holds not only during the slow--roll regime but also after reheating.
The reason that $\gnl\sim\taunl$ regardless of subsequent evolution beyond slow--roll may be understood if we split the contributions to the non--linear parameters into instrinsic terms, which depend on the instrinsic non--Gaussianity of $\delta\vp_I$ at late times, and gauge terms which do not. This could be seen in the moment transport techniques developed by Mulryne et.al.~\cite{Mulryne2011Moment}, where $\zeta$ is evaluated by evolving the field correlation functions from horizon--crossing to the time of interest, then gauge--transforming to $\zeta$ on uniform energy hypersurface. For model examples, see~\cite{Anderson:2012em,Seery:2012vj}. In particular, one can see that the second terms in the moment transport expressions Eqs.~(61) and (62) in~\cite{Anderson:2012em} for $\taunl$ and $\gnl$ would be of the same form up to some numerical factor of order $O(1)$ if they are dominated by one of the field bispectrum contributions, i.e. $\left\langle\delta\vp\delta\vp\delta\vp\right\rangle$. Thus it would be expected that the consistency relation $\gnl\sim\taunl$ should hold as long as the second terms both dominate in the full expressions for $\taunl$ and $\gnl$, even though $\taunl$ and $\gnl$ may still evolve in time. We intend to return to this in the future.

\subsection{Comments on $\gnl$}

So far for all two--field models considered in the lierature, $\gnl$ is at most of the same order of magnitude as $\taunl$ and is much less than the current observational limit in CMB experiments and large scale surveys which is about $O(10^5)$. It has been already shown recently by Elliston et.al.~\cite{Elliston2012Large} using the analytic slow--roll expressions for $\gnl$ in separable models that it is hard to engineer a model where $\gnl$ can be as large as $O(10^5)$ during inflation and dominates the statistics in the trispectrum, even if one goes beyond quadratic order in the potential. The reasons for this are summarised as follows: 

Looking at the analytic expression for $\gnl$, given in Eqs~(3.10) and (3.12) of~\cite{Elliston2012Large}, all the terms are multiplied by second order slow--roll parameters which are of order $O(10^{-4})$ in general. In order to have large $\gnl$, we need the prefactors $\tau_i$ and $g_i$ to be much larger than $10$, in particular $>O(10^7)$ if we want $\gnl\sim O(10^5)$. Yet some of the prefactors $\tau_i$ and $g_i$ are bounded from above by $10$, and even for those which are not, it requires extreme excessive fine--tuning for them to be of order $>O(10^8)$ compared to the conditions required for having large observable $\fnl$ and $\taunl$. Moreover, the region of parameter space where that extreme fine--tuned conditions can be realized in general coincides with those where quantum fluctuations become important over the classical drift of potential flow. As a result, it is difficult to engineer a model where $\gnl\sim O(10^5)$ during slow--roll inflation in multifield models. 

It remains to be seen beyond the slow--roll regime though. In particular, $\gnl$ could be dramatically enhanced such that it is above the observational limit after reheating. Yet we found the same conclusion here even with reheating taken into account for all the models we study. In some cases, $\gnl$ does increase dramatically from $0$ to $O(100)$ for some decay rates, for instance see Fig.~\ref{fig:tauNL_axion} for the two minima model. It may be expected that a larger hierarchy between the decay rates may thus produce a large observable $\gnl$. However this is beyond the current numerical capabilities of our code. 

\subsection{Relation between $\nfnl$ and $\ntaunl$}

The consistency relation Eq.~(\ref{eq:running_2field_local}) follows from the class of two--field local type models with $\zeta$ of the form \cite{Byrnes:2010ft}
\bea
\zeta(k) = \zeta_{k}^{G,\vp} + \zeta_{k}^{G,\chi} + f_{\vp}(\zeta^{G,\vp}\star\zeta^{G,\vp})_k + g_{\vp}(\zeta^{G,\vp}\star\zeta^{G,\vp}\star\zeta^{G,\vp})_k\,, \label{eq:2field_local}
\eea
when $f_\vp$ and $g_\vp$ are scale independent and $\zeta^{G,\vp}$, $\zeta^{G,\chi}$ are Gaussian variables \footnote{But not necessarily the opposite, i.e. the consistency relation Eq.~(\ref{eq:running_2field_local}) does not necessarily imply the model is of two--field local type.}. For all cases we study only one of the fields develops significant non-Gaussianity, so $\zeta$ may fit this ansatz. The question is whether $f_\vp$ and $g_\vp$ are scale independent for the models we study. They are if the field which generates non-Gaussianity is strongly subdominant, has a quadratic potential and no interactions with the inflaton field. Many of the models we study are approximately of this type, and hence we observe $3\nfnl\simeq2\ntaunl$.

For single source models there is a different consistency relation, which trivially follows from $\taunl\propto\fnl^2$,
\bea
\ntaunl = 2\nfnl\,. \label{eq:singlefield_runnings}
\eea
In the limit that $\zeta_{k}^{G,\vp} \ll \zeta_{k}^{G,\chi}$, which corresponds to $\Nvp^2\ll\Nchi^2\gg 1$, the model becomes effectively single source. If the assumptions related to (\ref{eq:2field_local}) remain valid, the non-linearity parameters have to be scale independent.

It is worth noting that $\zeta$ does not always satisfy the ansatz Eq.~(\ref{eq:2field_local}) in the models we study. For instance, for the two minima model, the relation $3\nfnl\simeq2\ntaunl$ only holds after the subdominate $\vp$ field starts oscillating but not during inflation, as shown in Fig.~\ref{fig:gNL_tauNL_axion} and Table~\ref{tab:twominstats}. As mentioned above, naively taking the predictions evaluated at the end of inflation, one would find $\ntaunl\ll\nfnl$ and therefore conclude that the model does not belong to the class of two--field local type models, which is clearly invalid when reheating is taken into account. Besides, reheating leads to non--trivial evolution of $\zeta$ and thus it is not obvious that given $\zeta$ satisfies Eq.~(\ref{eq:2field_local}) during inflation, this would continue to hold after reheating as shown in the one minimum case.

\section{Conclusions} \label{sec:conclus}

We have studied the evolution of the curvature perturbation through reheating, for the first time going up to third order in perturbation theory. This allows us to study the evolution of several observables during this period for the first time, namely the trispectrum consisting of two non-linearity parameters, $\gnl$ and $\taunl$ as well as the scale dependence of $\fnl$ and $\taunl$. The calculation during reheating is complex and requires numerical techniques, which to date has led to this field being rather neglected. However it is clearly very important, since reheating is required after inflation and observables will often not have reached their final value during inflation. It is of course only the final value which we may compare to observations, the evolution before is unobservable. In our reheating model, in which all scalar fields decay into radiation, the isocurvature mode will necessarily decay during this time and the curvature perturbation is thereafter conserved.

Of course the isocurvature mode does not have to decay during reheating in all models, for example it could be sustained by giving the inflaton fields multiple decay channels. But allowing them to decay into both matter and radiation alone does not appear to stop them decaying quickly \cite{Huston:2013kgl}. Even if $\zeta$ did evolve after reheating, our work is not redundant, however one would have to continue calculating the evolution until a later time \footnote{Note that in addition to the uncertainty of how observables are influenced by reheating, there is also an intrinsic uncertainty between the predicted global values of observables, and those which we measure in our Hubble volume. This effect is especially strong in models with local non-Gaussianity, due to the coupling between long and short wavelength modes~\cite{Nelson:2012sb,Nurmi:2013xv,LoVerde:2013xka}.}.

As we have found in our previous work~\cite{Leung:2012ve} for the case of $\fnl$ (see also~\cite{elliston:2011}), we find the trispectrum will in general be sensitive to the decay rates during reheating, although in some cases in which both fields oscillate after inflation, the sensitivity to the decay rates can be very small provided that the decay rates are equal. In general the evolution during reheating is large enough that a comparison of observable values between end of inflation and the final time would lead to the wrong conclusions, since the change in observables may be larger than the expected error bars of the observables. While the evolution to an adiabatic attractor during inflation often (but by no means always) results in a small value of $\fnl$ \cite{elliston:2011,Elliston2012Large,Meyers2011Adiabaticity}, this is not the case during reheating. Typically a model which is non-Gaussian at the end of inflation will remain non-Gaussian, and in most cases which we studied, the sign of the non-linearity parameters will also remain the same. The reverse is not always true, we have seen how in the axion model the perturbations are Gaussian at the end of inflation but not at the end of reheating. It would be interesting to study how generic these conclusions about the survival of non-Gaussianity are. It would also be interesting to study more realistic models of reheating and to include a period of non-perturbative preheating. However such studies are very difficult even following from single field inflation, and typically require a lattice simulation, which goes beyond the scope of this work.

Despite the evolution of all observables during reheating, we may still hope to test models of inflation against the new observational data. We have previously shown that typically the spectral index, $\nz-1$ is a more robust observable than non-Gaussianity, since it tends to be less sensitive to the details of reheating. With non-Gaussianity, we may instead look for consistency relations between the five observables which we have studied. First of all we have studied the well known Suyama-Yamaguchi inequality, $\taunl\geq(6\fnl/5)^2$, more specifically how strongly the equality may be broken. We have found that it remains very hard to generate $\taunl\gg\fnl^2$, consistent with other studies. We have provided some analytical and quite general insight into why this is the case. Interestingly, we have also found that for the two field inflation cases in which \cite{Elliston2012Large} found $\gnl\simeq\taunl$, that this relation typically remains true during reheating. Given the observational bounds on $\taunl$, it will be hard to observe $\gnl$ in such models. In fact we have not found any models with very large $\gnl$, despite studying several examples of models with a strong self interaction, so that their potential is far from quadratic. Finally we have also observed the relation between $\nfnl$ and $\ntaunl$, showing that in many cases $3\nfnl\simeq2\ntaunl$ both during and after inflation. These relations between observables allow the underlying model to be tested even when one cannot predict the actual values of any of the individual parameters.

%-------------------------------------------------------------------------------------------------------------
\acknowledgements

The authors would like to thank David Mulryne for useful discussions. The authors would also like to thank the organisers of the UKCOSMO meeting held at Imperial College London in March 2013 where part of the work was completed. GL and ERMT are supported by the University of Nottingham. CB is supported by a Royal Society University Research Fellowship. EJC acknowledges the STFC, Royal Society and Leverhulme Trust for financial support.

%-------------------------------------------------------------------------------------------------------------
\bibliographystyle{h-physrev}     
\bibliography{ppxet}

\begin{thebibliography}{10}

\bibitem{Guth1981Inflationary}
A.~H. Guth,
\newblock Physical Review D {\bf 23}, 347 (1981).

\bibitem{Linde1982New}
A.~Linde,
\newblock Physics Letters B {\bf 108}, 389 (1982).

\bibitem{Lyth2009Primordial}
D.~H. Lyth and A.~R. Liddle,
\newblock {\em {The Primordial Density Perturbation: Cosmology, Inflation and
  the Origin of Structure}} (Cambridge University Press, Cambridge, 2009).

\bibitem{Maldacena2003NonGaussian}
J.~Maldacena,
\newblock Journal of High Energy Physics {\bf 2003}, 013 (2003),
  astro-ph/0210603.

\bibitem{Seery2005Primordial}
D.~Seery and J.~E. Lidsey,
\newblock Journal of Cosmology and Astroparticle Physics {\bf 2005}, 011
  (2005), astro-ph/0506056.

\bibitem{Chen2007Large}
X.~Chen, R.~Easther, and E.~A. Lim,
\newblock Journal of Cosmology and Astroparticle Physics {\bf 2007}, 023
  (2007), astro-ph/0611645.

\bibitem{Chambers2010NonGaussianity}
A.~Chambers, S.~Nurmi, and A.~Rajantie,
\newblock Journal of Cosmology and Astroparticle Physics {\bf 2010}, 012
  (2010), 0909.4535.

\bibitem{Mollerach1990Isocurvature}
S.~Mollerach,
\newblock Physical Review D {\bf 42}, 313 (1990).

\bibitem{Lyth2002Generating}
D.~H. Lyth and D.~Wands,
\newblock Physics Letters B {\bf 524}, 5 (2002), hep-ph/0110002.

\bibitem{Linde2006Curvaton}
A.~Linde and V.~Mukhanov,
\newblock Journal of Cosmology and Astroparticle Physics {\bf 2006}, 009
  (2006), astro-ph/0511736.

\bibitem{Malik2006Numerical}
K.~A. Malik and D.~H. Lyth,
\newblock Journal of Cosmology and Astroparticle Physics {\bf 2006}, 008
  (2006), astro-ph/0604387.

\bibitem{Chambers2008Lattice}
A.~Chambers and A.~Rajantie,
\newblock Physical Review Letters {\bf 100} (2008), 0710.4133.

\bibitem{Chambers2008NonGaussianity}
A.~Chambers and A.~Rajantie,
\newblock Journal of Cosmology and Astroparticle Physics {\bf 2008}, 002+
  (2008), 0805.4795.

\bibitem{Kofman2003Probing}
L.~Kofman,
\newblock (2003), astro-ph/0303614.

\bibitem{Dvali2004New}
G.~Dvali, A.~Gruzinov, and M.~Zaldarriaga,
\newblock Physical Review D {\bf 69} (2004), astro-ph/0303591.

\bibitem{Suyama2008NonGaussianity}
T.~Suyama and M.~Yamaguchi,
\newblock Physical Review D {\bf 77} (2008), 0709.2545.

\bibitem{Byrnes2009Constraints}
C.~T. Byrnes,
\newblock Journal of Cosmology and Astroparticle Physics {\bf 2009}, 011
  (2009), 0810.3913.

\bibitem{Lyth2005Generating}
D.~H. Lyth,
\newblock Journal of Cosmology and Astroparticle Physics {\bf 2005}, 006
  (2005), astro-ph/0510443.

\bibitem{Bernardeau2002NonGaussianity}
F.~Bernardeau and J.-P. Uzan,
\newblock Physical Review D {\bf 66} (2002), hep-ph/0207295.

\bibitem{Alabidi:2006hg}
L.~Alabidi,
\newblock Journal of Cosmology and Astroparticle Physics {\bf 0610}, 015
  (2006), astro-ph/0604611.

\bibitem{Byrnes2009Large}
C.~T. Byrnes, K.-Y. Choi, and L.~M. Hall,
\newblock Journal of Cosmology and Astroparticle Physics {\bf 2009}, 017
  (2009), 0812.0807.

\bibitem{Byrnes2008Conditions}
C.~T. Byrnes, K.-Y. Choi, and L.~M.~H. Hall,
\newblock Journal of Cosmology and Astroparticle Physics {\bf 2008}, 008+
  (2008), 0807.1101.

\bibitem{Gao:2013hn}
X.~Gao and P.~Shukla,
\newblock (2013), 1301.6076.

\bibitem{Byrnes2010Review}
C.~T. Byrnes and K.-Y. Choi,
\newblock Advances in Astronomy {\bf 2010}, 1 (2010), 1002.3110.

\bibitem{Chen:2010xka}
X.~Chen,
\newblock Adv.Astron. {\bf 2010}, 638979 (2010), 1002.1416.

\bibitem{elliston:2011}
J.~Elliston, D.~J. Mulryne, D.~Seery, and R.~Tavakol,
\newblock (2011), 1106.2153.

\bibitem{Leung:2012ve}
G.~Leung, E.~R. Tarrant, C.~T. Byrnes, and E.~J. Copeland,
\newblock Journal of Cosmology and Astroparticle Physics {\bf 1209}, 008
  (2012), 1206.5196.

\bibitem{Choi2012Primordial}
K.-Y. Choi, S.~A. Kim, and B.~Kyae,
\newblock Nuclear Physics B {\bf 861}, 271 (2012), 1202.0089.

\bibitem{Sasaki2008Multibrid}
M.~Sasaki,
\newblock Progress of Theoretical Physics {\bf 120}, 159 (2008), 0805.0974.

\bibitem{Naruko2009Large}
A.~Naruko and M.~Sasaki,
\newblock Progress of Theoretical Physics {\bf 121}, 193 (2009), 0807.0180.

\bibitem{Peterson:2010mv}
C.~M. Peterson and M.~Tegmark,
\newblock Phys.Rev. {\bf D84}, 023520 (2011), 1011.6675.

\bibitem{Elliston2012Large}
J.~Elliston, L.~Alabidi, I.~Huston, D.~Mulryne, and R.~Tavakol,
\newblock Journal of Cosmology and Astroparticle Physics {\bf 2012}, 001
  (2012), 1203.6844.

\bibitem{Kofman1996Origin}
L.~Kofman,
\newblock (1996), astro-ph/9605155.

\bibitem{Huston:2013kgl}
I.~Huston and A.~J. Christopherson,
\newblock (2013), 1302.4298.

\bibitem{Bennett:2012fp}
C.~Bennett {\em et~al.},
\newblock (2012), 1212.5225.

\bibitem{Giannantonio:2013uqa}
T.~Giannantonio {\em et~al.},
\newblock (2013), 1303.1349.

\bibitem{Smidt:2010sv}
J.~Smidt {\em et~al.},
\newblock (2010), 1001.5026.

\bibitem{Smidt:2010ra}
J.~Smidt {\em et~al.},
\newblock Phys.Rev. {\bf D81}, 123007 (2010), 1004.1409.

\bibitem{Fergusson:2010gn}
J.~Fergusson, D.~Regan, and E.~Shellard,
\newblock (2010), 1012.6039.

\bibitem{Komatsu:2001rj}
E.~Komatsu and D.~N. Spergel,
\newblock Phys.Rev. {\bf D63}, 063002 (2001), astro-ph/0005036.

\bibitem{Kogo:2006kh}
N.~Kogo and E.~Komatsu,
\newblock Phys.Rev. {\bf D73}, 083007 (2006), astro-ph/0602099.

\bibitem{Chen:2005fe}
X.~Chen,
\newblock Physical Review {\bf D72}, 123518 (2005), astro-ph/0507053.

\bibitem{Byrnes:2010ft}
C.~T. Byrnes, M.~Gerstenlauer, S.~Nurmi, G.~Tasinato, and D.~Wands,
\newblock Journal of Cosmology and Astroparticle Physics {\bf 1010}, 004
  (2010), 1007.4277.

\bibitem{Shandera:2010ei}
S.~Shandera, N.~Dalal, and D.~Huterer,
\newblock Journal of Cosmology and Astroparticle Physics {\bf 1103}, 017
  (2011), 1010.3722.

\bibitem{Byrnes:2011gh}
C.~T. Byrnes, K.~Enqvist, S.~Nurmi, and T.~Takahashi,
\newblock Journal of Cosmology and Astroparticle Physics {\bf 1111}, 011
  (2011), 1108.2708.

\bibitem{Kobayashi:2012ba}
T.~Kobayashi and T.~Takahashi,
\newblock Journal of Cosmology and Astroparticle Physics {\bf 1206}, 004
  (2012), 1203.3011.

\bibitem{Sefusatti:2009xu}
E.~Sefusatti, M.~Liguori, A.~P. Yadav, M.~G. Jackson, and E.~Pajer,
\newblock Journal of Cosmology and Astroparticle Physics {\bf 0912}, 022
  (2009), 0906.0232.

\bibitem{Biagetti:2013sr}
M.~Biagetti, H.~Perrier, A.~Riotto, and V.~Desjacques,
\newblock (2013), 1301.2771.

\bibitem{Sasaki1996General}
M.~Sasaki and E.~D. Stewart,
\newblock Progress of Theoretical Physics {\bf 95}, 71 (1996),
  astro-ph/9507001.

\bibitem{Sasaki1998SuperHorizon}
M.~Sasaki and T.~Tanaka,
\newblock Progress of Theoretical Physics {\bf 99}, 763 (1998), gr-qc/9801017.

\bibitem{Lyth2005Inflationary}
D.~Lyth and Y.~Rodr\'{\i}guez,
\newblock Physical Review Letters {\bf 95} (2005), astro-ph/0504045.

\bibitem{Saffin2012Covariance}
P.~M. Saffin,
\newblock (2012), 1203.0397.

\bibitem{Elliston:2012ab}
J.~Elliston, D.~Seery, and R.~Tavakol,
\newblock Journal of Cosmology and Astroparticle Physics {\bf 1211}, 060
  (2012), 1208.6011.

\bibitem{Byrnes:2006vq}
C.~T. Byrnes, M.~Sasaki, and D.~Wands,
\newblock Physical Review {\bf D74}, 123519 (2006), astro-ph/0611075.

\bibitem{Byrnes:2009pe}
C.~T. Byrnes, S.~Nurmi, G.~Tasinato, and D.~Wands,
\newblock Journal of Cosmology and Astroparticle Physics {\bf 1002}, 034
  (2010), 0911.2780.

\bibitem{Lyth:1998xn}
D.~H. Lyth and A.~Riotto,
\newblock Phys.Rept. {\bf 314}, 1 (1999), hep-ph/9807278.

\bibitem{Dias2012Transport}
M.~Dias and D.~Seery,
\newblock Physical Review D {\bf 85} (2012), 1111.6544.

\bibitem{Huston2012Calculating}
I.~Huston and A.~Christopherson,
\newblock Physical Review D {\bf 85} (2012), 1111.6919.

\bibitem{Anderson:2012em}
G.~J. Anderson, D.~J. Mulryne, and D.~Seery,
\newblock Journal of Cosmology and Astroparticle Physics {\bf 1210}, 019
  (2012), 1205.0024.

\bibitem{Watanabe2012Delta}
Y.~Watanabe,
\newblock (2012), 1110.2462.

\bibitem{Peterson2011NonGaussianity}
C.~Peterson and M.~Tegmark,
\newblock Physical Review D {\bf 84} (2011), 1011.6675.

\bibitem{Mulryne2011Moment}
D.~J. Mulryne, D.~Seery, and D.~Wesley,
\newblock Journal of Cosmology and Astroparticle Physics {\bf 2011}, 030
  (2011), 1008.3159.

\bibitem{Vernizzi2006NonGaussianities}
F.~Vernizzi and D.~Wands,
\newblock Journal of Cosmology and Astroparticle Physics {\bf 2006}, 019
  (2006), astro-ph/0603799.

\bibitem{Suyama:2010uj}
T.~Suyama, T.~Takahashi, M.~Yamaguchi, and S.~Yokoyama,
\newblock Journal of Cosmology and Astroparticle Physics {\bf 1012}, 030
  (2010), 1009.1979.

\bibitem{Lewis:2011au}
A.~Lewis,
\newblock Journal of Cosmology and Astroparticle Physics {\bf 1110}, 026
  (2011), 1107.5431.

\bibitem{Smith:2011if}
K.~M. Smith, M.~LoVerde, and M.~Zaldarriaga,
\newblock Phys.Rev.Lett. {\bf 107}, 191301 (2011), 1108.1805.

\bibitem{Sugiyama:2012tr}
N.~S. Sugiyama,
\newblock Journal of Cosmology and Astroparticle Physics {\bf 1205}, 032
  (2012), 1201.4048.

\bibitem{Assassi:2012zq}
V.~Assassi, D.~Baumann, and D.~Green,
\newblock Journal of Cosmology and Astroparticle Physics {\bf 1211}, 047
  (2012), 1204.4207.

\bibitem{Kehagias:2012pd}
A.~Kehagias and A.~Riotto,
\newblock Nuclear Physics {\bf B864}, 492 (2012), 1205.1523.

\bibitem{Tasinato:2012js}
G.~Tasinato, C.~T. Byrnes, S.~Nurmi, and D.~Wands,
\newblock (2012), 1207.1772.

\bibitem{Rodriguez:2013cj}
Y.~Rodriguez, J.~P.~B. Almeida, and C.~A. Valenzuela-Toledo,
\newblock (2013), 1301.5843.

\bibitem{Seery:2012vj}
D.~Seery, D.~J. Mulryne, J.~Frazer, and R.~H. Ribeiro,
\newblock Journal of Cosmology and Astroparticle Physics {\bf 1209}, 010
  (2012), 1203.2635.

\bibitem{Nelson:2012sb}
E.~Nelson and S.~Shandera,
\newblock (2012), 1212.4550.

\bibitem{Nurmi:2013xv}
S.~Nurmi, C.~T. Byrnes, and G.~Tasinato,
\newblock (2013), 1301.3128.

\bibitem{LoVerde:2013xka}
M.~LoVerde, E.~Nelson, and S.~Shandera,
\newblock (2013), 1303.3549.

\bibitem{Meyers2011Adiabaticity}
J.~Meyers and N.~Sivanandam,
\newblock Physical Review D {\bf 84} (2011), 1104.5238.

\end{thebibliography}

\end{document}